\documentclass[sigconf]{acmart}
 
\pdfoutput=1

\def\BibTeX{{\rm B\kern-.05em{\sc i\kern-.025em b}\kern-.08emT\kern-.1667em\lower.7ex\hbox{E}\kern-.125emX}}

\usepackage{array}
\usepackage{balance}
\usepackage{graphicx}
\usepackage{clrscode}
\usepackage{subfigure}
\usepackage{multirow}
\usepackage{float}
\usepackage{color}
\usepackage{soul}
\usepackage{mdframed}
\usepackage{amsopn}
\usepackage{mathrsfs}
\usepackage{mathtools}
\usepackage{amsmath}
\usepackage{amsfonts}
\usepackage{arydshln}
\usepackage{hyperref}
\usepackage{multicol}
\usepackage{blkarray}
\usepackage{enumerate}
\usepackage{courier}
\usepackage{rotating}
\usepackage{booktabs}
\usepackage{diagbox}
\usepackage{fancybox}
\usepackage{minibox}
\usepackage{cases}
\usepackage[lined,boxruled,commentsnumbered,linesnumbered]{algorithm2e}
\usepackage{dsfont}
\usepackage{amsthm}

\newcommand{\etal}{\textit{et al}. }

\copyrightyear{2019} 
\acmYear{2019} 
\setcopyright{acmcopyright}
\acmConference[SIGIR '19]{Proceedings of the 42nd International ACM SIGIR Conference on Research and Development in Information Retrieval}{July 21--25, 2019}{Paris, France}
\acmBooktitle{Proceedings of the 42nd International ACM SIGIR Conference on Research and Development in Information Retrieval (SIGIR '19), July 21--25, 2019, Paris, France}
\acmPrice{15.00}
\acmDOI{10.1145/3331184.3331203}
\acmISBN{978-1-4503-6172-9/19/07}

\begin{document}

\fancyhead{}

\title{Reinforcement Knowledge Graph Reasoning for \\ Explainable Recommendation}

\author{Yikun Xian}
\authornote{Both authors contributed equally to this work.}
\affiliation{\institution{Rutgers University}}
\email{siriusxyk@gmail.com}

\author{Zuohui Fu}
\authornotemark[1]
\affiliation{\institution{Rutgers University}}
\email{zuohui.fu@rutgers.edu}

\author{S. Muthukrishnan}
\affiliation{\institution{Rutgers University}}
\email{muthu@cs.rutgers.edu}

\author{Gerard de Melo}
\affiliation{\institution{Rutgers University}}
\email{gdm@demelo.org}

\author{Yongfeng Zhang}
\affiliation{\institution{Rutgers University}}
\email{yongfeng.zhang@rutgers.edu}

\begin{abstract}
Recent advances in personalized recommendation have sparked great interest in the exploitation of rich structured information provided by knowledge graphs.
Unlike most existing approaches that only focus on leveraging knowledge graphs for more accurate recommendation, we perform explicit reasoning with knowledge for decision making so that the recommendations are generated and supported by an interpretable causal inference procedure.
To this end, we propose a method called Policy-Guided Path Reasoning (PGPR), which couples recommendation and interpretability by providing actual paths in a knowledge graph. Our contributions include four aspects. We first highlight the significance of incorporating knowledge graphs into recommendation to formally define and interpret the reasoning process. Second, we propose a reinforcement learning (RL) approach featuring an innovative soft reward strategy, user-conditional action pruning and a multi-hop scoring function. Third, we design a policy-guided graph search algorithm to efficiently and effectively sample reasoning paths for recommendation. Finally, we extensively evaluate our method on several large-scale real-world benchmark datasets, obtaining favorable results compared with state-of-the-art methods.
\end{abstract}



\maketitle

\section{Introduction}\label{sec:intro}

Equipping recommendation systems with the ability to leverage knowledge graphs (KG) not only facilitates better exploitation of various structured information to improve the recommendation performance, but also enhances the explainability of recommendation models due to the intuitive ease of understanding relationships between entities \cite{zhang2018explainable}. Recently, researchers have explored the potential of knowledge graph reasoning in personalized recommendation.
One line of research focuses on making recommendations using knowledge graph embedding models, such as TransE \cite{bordes2013translating} and node2vec \cite{grover2016node2vec}.
These approaches align the knowledge graph in a regularized vector space and uncover the similarity between entities by calculating their representation distance \cite{zhang2016collaborativekdd}. However, pure KG embedding methods lack the ability to discover multi-hop relational paths.
Ai \etal \cite{ai2018learning} proposed to enhance the collaborative filtering (CF) method over KG embedding for personalized recommendation, followed by a soft matching algorithm to find explanation paths between users and items.
However, one issue of this strategy is that the explanations are not produced according to the reasoning process, but instead are later generated by an empirical similarity matching between the user and item embeddings. Hence, their explanation component is merely trying to find a post-hoc explanation for the already chosen recommendations.

Another line of research investigates path-based recommendation.
For example, Gao \etal \cite{gao2018recommendation} proposed the notion of meta-paths to reason over KGs. However, the approach has difficulty in coping with numerous types of relations and entities in large real-world KGs, and hence it is incapable of exploring relationships between unconnected entities.
Wang \etal \cite{wang2018explainable} first developed a path embedding approach for recommendation over KGs that enumerates all the qualified paths between every user--item pair, and then trained a sequential RNN model from the extracted paths to predict the ranking score for the pairs. The recommendation performance is further improved, but it is not practical to fully explore all the paths for each user--item pair in large-scale KGs.

\begin{figure}[t]
\includegraphics[width=0.45\textwidth]{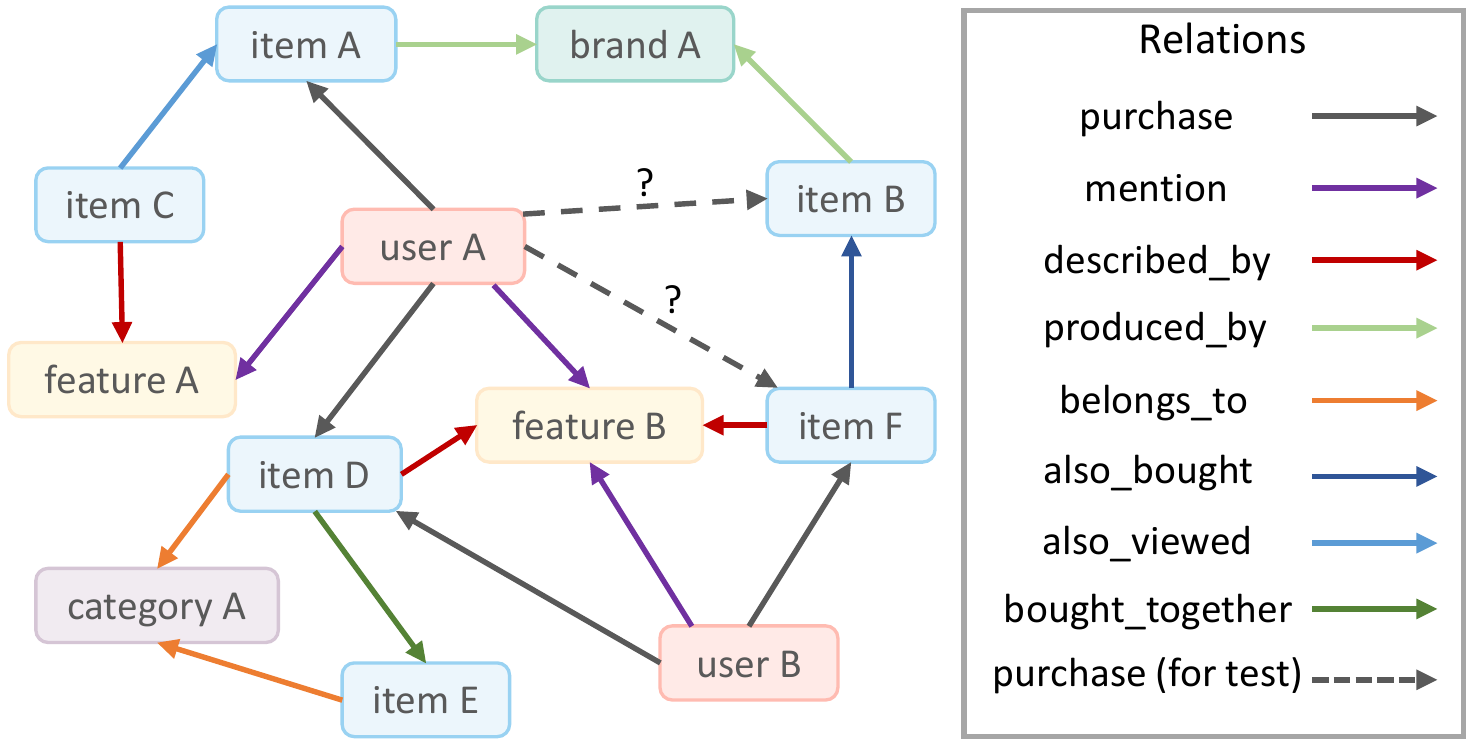}
\vspace{-10pt}
\caption{Illustration of the Knowledge Graph Reasoning for Explainable Recommendation (KGRE-Rec) problem.}
\label{fig:problem}
\vspace{-10pt}
\end{figure}

We believe that an intelligent recommendation agent should have the ability to conduct explicit reasoning over knowledge graphs to make decisions, rather than merely embed the graph as latent vectors for similarity matching.
In this paper, we consider knowledge graphs as a versatile structure to maintain the agent's knowledge about users, items, other entities and their relationships. The agent starts from a user and conducts explicit multi-step path reasoning over the graph, so as to discover suitable items in the graph for recommendation to the target user.
The underlying idea is that if the agent draws its conclusion based on an explicit reasoning path, it will be easy to interpret the reasoning process that leads to each recommendation. Thus, the system can provide causal evidence in support of the recommended items. Accordingly, our goal is not only to select a set of candidate items for recommendation, but also to provide the corresponding reasoning paths in the graph as interpretable evidence for why a given recommendation is made.
As an example illustrated in Figure \ref{fig:problem}, given user $A$, the algorithm is expected to find candidate items $B$ and $F$, along with their reasoning paths in the graph, e.g., $\{\text{User }A ~\rightarrow~ \text{Item }A ~\rightarrow~ \text{Brand }A ~\rightarrow~ \text{Item }B\}$ and $\{\text{User }A ~\rightarrow~ \text{Feature }B ~\rightarrow~ \text{Item }F\}$.

In this paper, we propose an approach that overcomes the shortcomings of previous work.
Specifically, we cast the recommendation problem as a deterministic Markov Decision Process (MDP) over the knowledge graph. We adopt a Reinforcement Learning (RL) approach, in which an agent starts from a given user, and learns to navigate to the potential items of interest, such that the path history can serve as a genuine explanation for why the item is recommended to the user.

The main challenges are threefold. 
First, it is non-trivial to measure the correctness of an item for a user, so careful consideration is needed regarding the terminal conditions and RL rewards. To solve the problem, we design a soft reward strategy based on a multi-hop scoring function that leverages the rich heterogeneous information in the knowledge graph.
Second, the size of the action space depends on the out-degrees in the graph, which can be very large for some nodes, so it is important to conduct an efficient exploration to find promising reasoning paths in the graph. In this regard, we propose a user-conditional action pruning strategy to decrease the size of the action spaces while guaranteeing the recommendation performance. 
Third, the diversity of both items and paths must be preserved when the agent is exploring the graph for recommendation, so as to avoid being trapped in limited regions of items. To achieve this, we design a policy-guided search algorithm to sample reasoning paths for recommendation in the inference phase. We conduct several case studies on the reasoning paths to qualitatively evaluate the diversity of explainations for recommendation.

The major contributions of this paper can be outlined as follows.
\begin{enumerate}
\item We highlight the significance of incorporating rich heterogeneous information into the recommendation problem to formally define and interpret the reasoning process.
\item We propose an RL-based approach to solve the problem, driven by our soft reward strategy, user-conditional action pruning, and a multi-hop scoring strategy.
\item We design a beam search-based algorithm guided by the policy network to efficiently sample diverse reasoning paths and candidate item sets for recommendation.
\item We extensively evaluate the effectiveness of our method on several Amazon e-commerce domains, obtaining strong results as well as explainable reasoning paths.
\end{enumerate}
The source code is available online.\footnote{Link to the source code: \url{https://github.com/orcax/PGPR}}

\section{Related Work} \label{sec:related}

\subsection{Collaborative Filtering}
Collaborative Filtering (CF) has been one of the most fundamental approaches for recommendation. 
Early approaches to CF consider the user--item rating matrix and predict ratings via user-based \cite{resnick1994grouplens,konstan1997grouplens} or item-based \cite{sarwar2001item,linden2003amazon} collaborative filtering methods. 
With the development of dimension reduction methods, latent factor models such as matrix factorization gained widespread adoption in recommender systems. Specific techniques include singular value decomposition \cite{koren2009matrix}, non-negative matrix factorization \cite{lee2001algorithms} and probabilistic matrix factorization \cite{mnih2008probabilistic}. For each user and item, these approaches essentially learn a latent factor representation to calculate the matching score of the user--item pairs.
Recently, deep learning and neural models have further extended collaborative filtering. These are broadly classified into two sub-categories: the similarity learning approach and the representation learning approach. Similarity learning adopts fairly simple user/item embeddings (e.g., one-hot vectors) and learns a complex prediction network as a similarity function to compute user--item matching scores \cite{he2017neural}. In contrast, the representation learning approach learns much richer user/item representations but adopts a simple similarity function (e.g., inner product) for score matching \cite{zhang2017joint}. However, researchers have noticed the difficulty of explaining the recommendation results in latent factor or latent representation models, making explainable recommendation \cite{explain,zhang2018explainable} an important research problem for the community.

\begin{figure*}[t]
    \centering
    \includegraphics[width=0.95\textwidth]{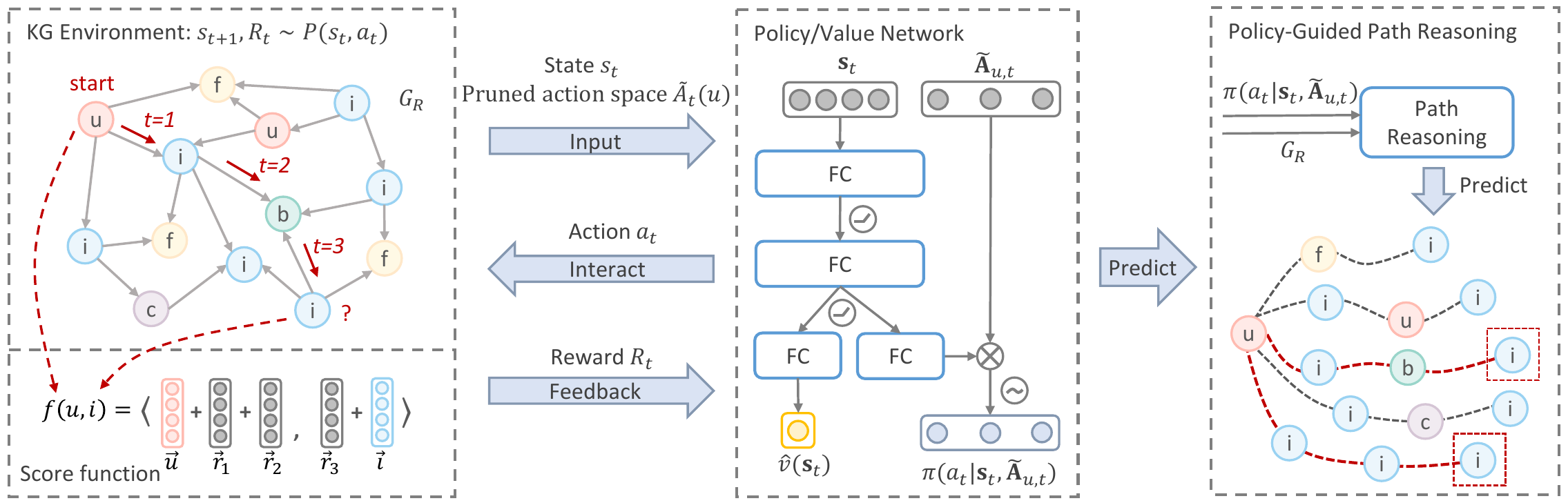}
    \vspace{-10pt}
    \caption{Pipeline of our Policy-Guided Path Reasoning method for recommendation. The algorithm aims to learn a policy that navigates from a user to potential items of interest by interacting with the knowledge graph environment. The trained policy is then adopted for the path reasoning phase to make recommendations to the user.}
    \label{fig:method}
    \vspace{-10pt}
\end{figure*}

\subsection{Recommendation with Knowledge Graphs}
Some previous efforts have made recommendations to users with the help of knowledge graph embeddings \cite{nickel2011three,bordes2013translating}. One research direction leverages knowledge graph embeddings as rich content information to enhance the recommendation performance. For example,
Zhang \etal \cite{zhang2016collaborativekdd} adopted knowledge base embeddings to generate user and item representations for recommendation, while Huang \etal \cite{huang2018improving} employed memory networks over knowledge graph entity embeddings for recommendation. Wang \etal \cite{wang2018ripplenet} proposed a ripple network approach for embedding-guided multi-hop KG-based recommendation. Another research direction attempts to leverage the entity and path information in the knowledge graph to make explainable decisions.
For example, Ai \etal~\cite{ai2018learning} incorporated the learning of knowledge graph embeddings for explainable recommendation. However, their explanation paths are essentially post-hoc explanations, as they are generated by soft matching after the corresponding items have been chosen.
Wang \etal \cite{wang2018explainable} proposed an RNN based model to reason over KGs for recommendation. However, it requires enumerating all the possible paths between each user--item pair for model training and prediction, which can be impractical for large-scale knowledge graphs.

\subsection{Reinforcement Learning}
Reinforcement Learning has attracted substantial interest in the research community. In recent years, there have been a series of widely noted successful applications of deep RL approaches (e.g., AlphaGo \cite{silver2016mastering}), demonstrating their ability to better understand the environment, and enabling them to infer high-level causal relationships.
There have been attempts to invoke RL in recommender systems in a non-KG setting, such as for ads recommendation \cite{theocharous2015personalized}, news recommendation \cite{zheng2018drn} and post-hoc explainable recommendation \cite{wang2018reinforcement}. %
At the same time, researchers have also explored RL in KG settings for other tasks such as question answering (QA) \cite{xiong2017deeppath,das2017go,lin2018multi}, which formulates multi-hop reasoning as a sequential decision making problem. 
For example, Xiong \etal \cite{xiong2017deeppath} leveraged reinforcement learning for path-finding, and Das \etal \cite{das2017go} proposed a system called MINERVA that trains a model for multi-hop KG question answering.
Lin \etal \cite{lin2018multi} proposed models for end-to-end RL-based KG question answering with reward shaping and action dropout. However, to the best of our knowledge, there is no previous work utilizing RL in KGs for the task of recommendation, especially when the KG has an extremely large action space for each entity node as the number of path hops grow.

\section{Methodology}\label{sec:method}
In this section, we first formalize a new recommendation problem called \emph{Knowledge Graph Reasoning for Explainable Recommendation}. 
Then we present our approach based on reinforcement learning over knowledge graphs to solve the problem.

\subsection{Problem Formulation}
In general, a knowledge graph $\mathcal{G}$ with entity set $\mathcal{E}$ and relation set $\mathcal{R}$ is defined as $\mathcal{G}=\{(e, r, e') \mid e,e'\in\mathcal{E}, r\in\mathcal{R}\}$, where each triplet $(e,r,e')$ represents a fact of the relation $r$ from head entity $e$ to tail entity $e'$.
In this paper, we consider a special type of knowledge graph for explainable recommendation, denoted by $\mathcal{G}_\text{R}$. 
It contains a subset of a \emph{User} entities $\mathcal{U}$ and a subset of \emph{Item} entities $\mathcal{I}$, where $\mathcal{U},\mathcal{I}\subseteq\mathcal{E}$ and $\mathcal{U}\cap\mathcal{I}=\varnothing$.
These two kinds of entities are connected through relations $r_{ui}$.
We give a relaxed definition of $k$-hop paths over the graph $\mathcal{G}_\text{R}$ as follows.

\begin{definition}{($k$-hop path)}
A $k$-hop path from entity $e_0$ to entity $e_k$ is defined as a sequence of $k+1$ entities connected by $k$ relations, denoted by $p_k(e_0,e_k)=\left\{e_0 \xleftrightarrow{r_1} e_1 \xleftrightarrow{r_2} \cdots \xleftrightarrow{r_k} e_k \right\}$, where $e_{i-1}\xleftrightarrow{r_i} e_{i}$ represents either $(e_{i-1},r_i,e_i)\in\mathcal{G}_\text{R}$ or $(e_{i},r_i,e_{i-1})\in\mathcal{G}_\text{R}$, $i\in [k]$.
\end{definition}

Now, the problem of \emph{Knowledge Graph Reasoning for Explainable Recommendation (KGRE-Rec)} can be formalized as below.

\begin{definition}{(KGRE-Rec Problem)}
Given a knowledge graph $\mathcal{G}_\text{R}$, user $u\in\mathcal{U}$ and integers $K$ and $N$, the goal is to find a recommendation set of items $\{i_n\}_{n\in[N]}\subseteq\mathcal{I}$ such that each pair $(u,i_n)$ is associated with one reasoning path $p_k(u,i_n)~(2\le k\le K)$, and $N$ is the number of recommendations.
\end{definition}

In order to simultaneously conduct item recommendation and path finding, we consider three aspects that result in a good solution to the problem.
First, we do not have pre-defined targeted items for any user, so it is not applicable to use a binary reward indicating whether the user interacts with the item or not. A better design of the reward function is to incorporate the uncertainty of how an item is relevant to a user based on the rich heterogeneous information given by the knowledge graph.
Second, out-degrees of some entities may be very large, which degrades the efficiency of finding paths from users to potential item entities.
Enumeration of all possible paths between each user and all items is unfeasible on very large graphs.
Thus, the key challenge is how to effectively perform edge pruning and efficiently search relevant paths towards potential items using the reward as a heuristic.
Third, for every user, the diversity of reasoning paths for recommended items should be guaranteed. It is not reasonable to always stick to a specific type of reasoning path to provide explainable recommendations.
One naive solution is post-hoc recommendation, which first generates candidate items according to some similarity measure, followed by a separate path finding procedure from the user to candidate items within the graph.
The major downsides of this are that the recommendation process fails to leverage the rich heterogeneous meta-data in the knowledge graph, 
and that the generated paths are detached from the actual decision-making process adopted by the recommendation algorithm, which remains uninterpretable.

In the following sections, we introduce our \emph{Policy-Guided Path Reasoning} method (PGPR) for explainable recommendation over knowledge graphs.
It solves the problem through reinforcement learning by making recommendations while simultaneously searching for paths in the context of rich heterogeneous information in the KG. As illustrated in Figure \ref{fig:method}, the main idea is to train an RL agent that learns to navigate to potentially ``good'' items conditioned on the starting user in the knowledge graph environment.
The agent is then exploited to efficiently sample reasoning paths for each user leading to the recommended items. These sampled paths naturally serve as the explanations for the recommended items. 

\subsection{Formulation as Markov Decision Process}
The starting point of our method is to formalize the KGRE-Rec problem as a Markov Decision Process (MDP) \cite{sutton2018reinforcement}.
In order to guarantee path connectivity, we add two special kinds of edges to the graph $\mathcal{G}_{\text{R}}$. The first one are reverse edges, i.e., if $(e,r,e')\in\mathcal{G}_{\text{R}}$, then $(e',r,e)\in\mathcal{G}_{\text{R}}$, which are used for our path definition. The second are self-loop edges, associated with the no operation (NO-OP) relation, i.e., if $e\in\mathcal{E}$, then $(e,r_\mathrm{noop},e)\in\mathcal{G}_{\text{R}}$.

\paragraph{State} 
The state $s_t$ at step $t$ is defined as a tuple $(u, e_t, h_t)$, where $u\in\mathcal{U}$ is the starting user entity, $e_t$ is the entity the agent has reached at step $t$, and $h_t$ is the history prior to step $t$. 
We define the $k$-step history as the combination of all entities and relations in the past $k$ steps, i.e., $\{e_{t-k},r_{t-k+1},\ldots,e_{t-1},r_t\}$.
Conditioned on some user $u$, the initial state is represented as $s_\text{0}=(u,u,\varnothing)$.
Given some fixed horizon $T$, the terminal state is $s_T=(u,e_T,h_T)$.

\paragraph{Action}
The complete action space $A_t$ of state $s_t$ is defined as all possible outgoing edges of entity $e_t$ excluding history entities and relations.
Formally, $A_t=\{(r,e) \mid (e_t,r,e)\in\mathcal{G}_{\text{R}}, e\not\in\{e_0,\ldots,e_{t-1}\}\}$. 
Since the out-degree follows a long-tail distribution, some nodes have much larger out-degrees compared with the rest of nodes.
It is fairly space-inefficient to maintain the size of the action space based on the largest out-degree. 
Thus, we introduce a \emph{user-conditional action pruning strategy} that effectively keeps the promising edges conditioned on the starting user based on a scoring function.
Specifically, the scoring function $f((r,e) \mid u)$ maps any edge $(r,e) ~(\forall r\in\mathcal{R}, \forall e \in\mathcal{E})$ to a real-valued score conditioned on user $u$.
Then, the user-conditional pruned action space of state $s_t$, denoted by $\tilde{A}_{t}(u)$, is defined as:
\begin{equation}\label{eq:act}
\tilde{A}_{t}(u)=\{(r,e) \mid \text{rank}(f((r,e) \mid u))\le\alpha,(r,e)\in A_t\},
\end{equation}
where $\alpha$ is a pre-defined integer that upper-bounds the size of the action space.
The details of this scoring function $f((r,e) \mid u)$ will be discussed in the next section.

\paragraph{Reward}
Given any user, there is no pre-known targeted item in the KGRE-Rec problem, so it is unfeasible to consider binary rewards indicating whether the agent has reached a target or not.
Instead, the agent is encouraged to explore as many ``good'' paths as possible.
Intuitively, in the context of recommendations, a ``good'' path is one that leads to an item that a user will interact with, with high probability.
To this end, we consider to give a soft reward only for the terminal state $s_T=(u,e_T,h_T)$ based on another scoring function $f(u,i)$. The terminal reward $R_T$ is defined as
\begin{equation}\label{eq:reward}
R_T = 
\begin{cases}
\max \left(0,\frac{f(u,e_T)}{\max_{i\in\mathcal{I}}f(u,i)}\right), & \text{~if~} e_T\in\mathcal{I} \\
0, & \text{~otherwise,}
\end{cases}
\end{equation}
where the value of $R_T$ is normalized to the range of $[0, 1]$. $f(u,i)$ is also introduced in the next section.

\paragraph{Transition}
Due to the graph properties, a state is determined by the position of the entity. 
Given a state $s_t=(u,e_t,h_t)$ and an action $a_t=(r_{t+1},e_{t+1})$, the transition to the next state $s_{t+1}$ is: 
\begin{equation}
\mathds{P}\left[s_{t+1}=(u,e_{t+1},h_{t+1})|s_t=(u,e_t,h_t),a_t=(r_{t+1},e_{t+1})\right]=1
\end{equation}
One exception is that the initial state $s_0=(u,u,\varnothing)$ is stochastic, which is determined by the starting user entity.
For simplicity, we assume the prior distribution of users follows a uniform distribution so that each user is equally sampled at the beginning.

\paragraph{Optimization}
Based on our MDP formulation, our goal is to learn a stochastic policy $\pi$ that maximizes the expected cumulative reward for any initial user $u$:
\begin{equation}
J(\theta)=\mathds{E}_{\pi}\left[ \left. \sum_{t=0}^{T-1}\gamma^{t}R_{t+1}\,\right\vert\, s_0=(u,u,\varnothing)\right].
\end{equation}
We solve the problem through \emph{REINFORCE with baseline} \cite{sutton2018reinforcement} by designing a policy network and a value network that share the same feature layers.
The policy network $\pi(\cdot|\mathbf{s},\mathbf{\tilde{A}}_u)$ takes as input the state vector $\mathbf{s}$ and binarized vector $\mathbf{\tilde{A}}_u$ of pruned action space $\tilde{A}(u)$ and emits the probability of each action, with zero probability for actions not in $\tilde{A}(u)$.
The value network $\hat{v}(\mathbf{s})$ maps the state vector $\mathbf{s}$ to a real value, which is used as the baseline in REINFORCE.
The structures of the two networks are defined as follows:
\begin{align}
 \mathbf{x}=&\,\text{dropout}(\sigma(\text{dropout}(\sigma(\mathbf{s} \mathbf{W}_1))\mathbf{W}_2)) \\
 \pi(\cdot|\mathbf{s},\mathbf{\tilde{A}}_u)=&\, \text{softmax}(\mathbf{\tilde{A}}_u\odot (\mathbf{x}\mathbf{W}_\text{p})) \\
 \hat{v}(\mathbf{s})=&\,\mathbf{x}\mathbf{W}_\text{v}
\end{align}
Here, $\mathbf{x}\in\mathds{R}^{d_f}$ are the learned hidden features of the state, $\odot$ is the Hadamard product, which is used to mask invalid actions here, and $\sigma$ is a non-linear activation function, for which we use an Exponential Linear Unit (ELU).
State vectors $\mathbf{s}\in\mathds{R}^{d_s}$ are represented as the concatenation of the embeddings $u$, $e_t$ and history ${h_t}$.
For the binarized pruned action space $\mathbf{\tilde{A}}_u\in\{0,1\}^{d_A}$, we set the maximum size $d_A$ among all pruned action spaces. 
The model parameters for both networks are denoted as $\Theta=\{\mathbf{W}_1, \mathbf{W}_2, \mathbf{W}_\text{p}, \mathbf{W}_\text{v}\}$.
Additionally, we add a regularization term $H(\pi)$ that maximizes the entropy of the policy in order to encourage the agent to explore more diverse paths.
Finally, the policy gradient $\nabla_{\Theta}J(\Theta)$ is defined as:  
\begin{equation}
\nabla_{\Theta} J(\Theta)=\mathds{E}_{\pi}\left[\nabla_{\Theta}\log \pi_{\Theta}(\cdot|\mathbf{s},\mathbf{\tilde{A}}_u) \left(G-\hat{v}(\mathbf{s})\right) \right],
\end{equation}
where $G$ is the discounted cumulative reward from state $s$ to the terminal state $s_T$.

\subsection{Multi-Hop Scoring Function}
Now we present the scoring function for the action pruning strategy and the reward function. We start with some relevant concepts.

One property of the knowledge graph $\mathcal{G}_\text{R}$ is that given the type of a head entity and a valid relation, the type of tail entity is determined. We can extend this property by creating a chain rule of entity and relation types: $\{e_0,r_1, e_1, r_2, \ldots, r_k, e_k\}$. If the types of entity $e_0$ and all relations $r_1,\ldots,r_k$ are given, the types of all other entities $e_1,\ldots,e_k$ are uniquely determined.
According to this rule, we introduce the concept of \emph{patterns} as follows.
\begin{definition}{($k$-hop pattern)}
A sequence of $k$ relations  $\tilde{r}_k=\{r_1,\ldots,r_k\}$ is called a valid $k$-hop pattern for two entities $(e_0,e_k)$ if there exists a set of entities $\{e_1,\ldots,e_{k-1}\}$ whose types are uniquely determined such that $\{e_0 \xleftrightarrow{r_1} e_1 \xleftrightarrow{r_2} \cdots \xleftrightarrow{r_{k-1}}e_{k-1}\xleftrightarrow{r_k} e_k \}$ forms a valid $k$-hop path over $\mathcal{G}_\text{R}$.
\end{definition}

One caveat with pattern is the direction of each relation, provided that we allow reverse edges in the path. For entities $e,e'$ and relation $r$, $e\xleftrightarrow{r} e'$ represents either $e\xrightarrow{r} e'$ or $e\xleftarrow{r} e'$ in the path.
We refer to the relation $r$ as a \emph{forward} one if $(e,r,e')\in\mathcal{G}_\text{R}$ and $e\xrightarrow{r} e'$, or as a \emph{backward} one if $(e',r,e)\in\mathcal{G}_\text{R}$ and $e\xleftarrow{r} e'$.

In order to define the scoring functions for action pruning and reward, we consider a special case of patterns with both forward and backward relations.
\begin{definition}{(\emph{1-reverse} $k$-hop pattern)}
A $k$-hop pattern is \emph{1-reverse}, denoted by $\tilde{r}_{k,j}=\{r_1,\ldots,r_j,r_{j+1},\ldots,r_k\}~(j\in[0,k])$, if $r_1,\ldots,r_j$ are forward and $r_{j+1}\ldots r_k$ are backward.
\end{definition}
In other words, paths with a \emph{1-reverse} $k$-hop pattern have the form of $e_0\xrightarrow{r_1}\cdots \xrightarrow{r_j}e_j\xleftarrow{r_{j+1}}e_{j+1}\xleftarrow{r_{j+2}}\cdots\xleftarrow{r_k} e_{k}$.
Note that the pattern contains all backward relations when $j=0$, and all forward relations when $j=k$.

Now we define a general multi-hop scoring function $f(e_0,e_k \mid \tilde{r}_{k,j})$ of two entities $e_0,e_k$ given \emph{1-reverse} $k$-hop pattern $\widetilde{r}_{k,j}$ as follows.
\begin{equation}\label{eq:score}
f(e_0,e_k \mid \widetilde{r}_{k,j}) = \left\langle\mathbf{e_0}+\sum_{s=1}^{j}\mathbf{r_s},~\mathbf{e_k}+\sum_{s=j+1}^{k}\mathbf{r_s}\right\rangle+b_{e_k},
\end{equation}
where $\langle\cdot,\cdot\rangle$ is the dot product operation, $\mathbf{e}$, $\mathbf{r}\in\mathds{R}^d$ are $d$-dimensional vector representations of the entity $e$ and relation $r$, and $b_{e}\in\mathds{R}$ is the bias for entity $e$.
When $k=0$, $j=0$, the scoring function simply computes the cosine similarity between two vectors:
\begin{equation}
f(e_0,e_k \mid \tilde{r}_{0,0})=\langle\mathbf{e_0},\mathbf{e_k}\rangle+b_{e_k}.
\end{equation}
When $k=1$,$j=1$, the scoring function computes the similarity between two entities via translational embeddings \cite{bordes2013translating}:
\begin{equation}\label{eq:transe}
f(e_0,e_k \mid \widetilde{r}_{1,1}) = \left\langle\mathbf{e_0}+\mathbf{r_1},~\mathbf{e_k}\right\rangle+b_{e_k}
\end{equation}
For $k\ge1,1\le j\le k$, the scoring function in Equation \ref{eq:score} quantifies the similarity of two entities based on a 1-reverse $k$-hop pattern.

\paragraph{Scoring Function for Action Pruning}
We assume that for user entity $u$ and another entity $e$ of other type, there exists only one \emph{1-reverse} $k$-hop pattern $\tilde{r}_{k,j}$ for some integer $k$.
For entity $e\not\in\mathcal{U}$, we denote $k_e$ as the smallest $k$ such that $\tilde{r}_{k,j}$ is a valid pattern for entities $(u,e)$.
Therefore, the scoring function for action pruning is defined as $f((r,e)\mid u)=f(u,e \mid \tilde{r}_{k_e,j})$.

\paragraph{Scoring Function for Reward}
We simply use the 1-hop pattern between user entity and item entity, i.e., $(u,r_{ui},i)\in\mathcal{G}_\text{R}$.
The scoring function for reward design is defined as $f(u,i)=f(u,i|\tilde{r}_{1,1})$.

\paragraph{Learning Scoring Function}
A natural question that arises is how to train the embeddings for each entity and relation. 
For any pair of entities $(e,e')$ with valid $k$-hop pattern $\tilde{r}_{k,j}$, we seek to maximize the conditional probability of $\mathds{P}(e' \mid e,\tilde{r}_{k,j})$, which is defined as
\begin{equation}
\mathds{P}(e'\mid e,\tilde{r}_{k,j})=\frac{\exp{\left( f(e,e'\mid\tilde{r}_{k,j}) \right)}}{\sum_{e''\in\mathcal{E}}\exp{\left( f(e,e''\mid\tilde{r}_{k,j}) \right)}}.
\end{equation}
However, due to the huge size of the entity set $\mathcal{E}$, we adopt a negative sampling technique \cite{mikolov2013distributed} to approximate $\log\mathds{P}(e'\mid e,\tilde{r}_{k,j})$:
\begin{align}
\log \mathds{P}(e|e',\tilde{r}_{k,j})\approx &\log\sigma\left(f(e,e'\mid \tilde{r}_{k,j})\right)  \nonumber\\
& +m\mathds{E}_{e''}\left[\log\sigma\left(-f(e,e''\mid\tilde{r}_{k,j})\right)\right]
\end{align}
The goal is to maximize the objective function $J(\mathcal{G}_\text{R})$, defined as:
\begin{equation}\label{eq:embed_obj}
J(\mathcal{G}_\text{R}) = \sum_{e,e'\in\mathcal{E}}\sum_{k=1}^{K}\mathds{1}\{(e,\tilde{r}_{k,j},e')\}\log\mathds{P}(e'|e,\tilde{r}_{k,j}),
\end{equation}
where $\mathds{1}\{(e,\tilde{r}_{k,j},e')\}$ is 1 if $\tilde{r}_{k,j}$ is a valid pattern for entities $(e,e')$ and 0 otherwise.

\subsection{Policy-Guided Path Reasoning}
\begin{algorithm}[t]
\caption{Policy-Guided Path Reasoning}
\label{alg:beam}
\SetKwInOut{Input}{Input}\SetKwInOut{Output}{Output}
\Input{$u$, $\pi(\cdot|\mathbf{s},\mathbf{\tilde{A}}_u)$, $T$, $\{K_1,\ldots,K_T\}$}
\Output{path set $\mathcal{P}_T$, probability set $\mathcal{Q}_T$, reward set $\mathcal{R}_T$}
\BlankLine
Initialize $\mathcal{P}_0\leftarrow\{\{u\}\}$, $\mathcal{Q}_0\leftarrow\{1\}$, $\mathcal{R}_0\leftarrow\{0\}$\;
\For{$t \leftarrow 1$ \KwTo $T$}{
  Initialize $\mathcal{P}_{t}\leftarrow\varnothing$, $\mathcal{Q}_t\leftarrow\varnothing$, $\mathcal{R}_t\leftarrow\varnothing$\;
  \ForAll{$\hat{p}\in \mathcal{P}_{t-1}$, $\hat{q}\in \mathcal{Q}_{t-1}$, $\hat{r}\in \mathcal{R}_{t-1}$}{
    \Comment{path $\hat{p}\doteq \{u,r_1,\ldots,r_{t-1},e_{t-1}\}$}\;
    Set $s_{t-1}\leftarrow (u,e_{t-1},h_{t-1})$\;
    Get user-conditional pruned action space $\tilde{A}_{t-1}(u)$ from environment given state $s_{t-1}$\;
    \Comment{$p(a)\doteq\pi(a\mid \mathbf{s}_{t-1},\mathbf{\tilde{A}}_{u,t-1})$ and $a\doteq (r_t,e_t)$}\;
    Actions $\mathcal{A}_t\leftarrow \left\{a\mid \text{rank}(p(a)) \le K_t,~\forall a\in \tilde{A}_{t-1}(u) \right\}$\;
    \ForAll{$a\in \mathcal{A}_t$}{
        Get $s_t, R_{t+1}$ from environment given action $a$\;
        Save new path $\hat{p}\cup\{r_t,e_t\}$ to $\mathcal{P}_{t}$\;
        Save new probability $p(a)\,\hat{q}$ to $\mathcal{Q}_t$\;
        Save new reward $R_{t+1}+\hat{r}$ to $\mathcal{R}_t$\;
    }
  }
}
Save $\forall\hat{p}\in \mathcal{P}_T$ if the path $\hat{p}$ ends with an item\;
\Return filtered $\mathcal{P}_T,\mathcal{Q}_T,\mathcal{R}_T$\;
\end{algorithm}

\begin{table*}[t]
\small
\centering
\begin{tabular}{cccccc}
\hline
& & \textbf{CDs \& Vinyl} & \textbf{Clothing} & \textbf{Cell Phones} & \textbf{Beauty} \\
\hline
\textbf{Entities} & \textbf{Description} & \multicolumn{4}{c}{\textbf{Number of Entities}} \vspace{0.2em} \\
\emph{User} & User in recommender system & \hphantom{0}75,258  & 39,387 & 27,879 & 22,363 \\
\emph{Item} & Product to be recommended to users & \hphantom{0}64,443  & 23,033 & 10,429 & 12,101 \\
\emph{Feature} & A product feature word from reviews & 202,959 & 21,366 & 22,493 & 22,564 \\
\emph{Brand} & Brand or manufacturer of the product & \hphantom{00}1,414   & \hphantom{0}1,182  & \hphantom{00,}955   & \hphantom{0}2,077 \\
\emph{Category} & Category of the product & \hphantom{000,}770     & \hphantom{0}1,193  & \hphantom{00,}206    & \hphantom{00,}248 \\
\hline
\textbf{Relations} & \textbf{Description} & \multicolumn{4}{c}{\textbf{Number of Relations per Head Entity}} \vspace{0.2em} \\
\emph{Purchase} & \emph{User} $~\xrightarrow{\text{purchase}}~$ \emph{Item} & $14.58\pm39.13$ &  $7.08\pm3.59$ & $6.97\pm4.55$ & $8.88\pm8.16$ \\
\emph{Mention} & \emph{User} $~\xrightarrow{\text{mention}}~$ \emph{Feature} & $\hphantom{0}2,545.92\pm10,942.31$ & $440.20\pm452.38$ & $\hphantom{0}652.08\pm1335.76$ & $\hphantom{0}806.89\pm1344.08$ \\
\emph{Described\_by} & \emph{Item} $~\xrightarrow{\text{described\_by}}~$ \emph{Feature} & $2,973.19\pm5,490.93$ & $752.75\pm909.42$ & $1,743.16\pm3,482.76$ & $1,491.16\pm2,553.93$ \\
\emph{Belong\_to} & \emph{Item} $~\xrightarrow{\text{belong\_to}}~$ \emph{Category} & $7.25\pm3.13$ & $6.72\pm2.15$ & $3.49\pm1.08$ & $4.11\pm0.70$ \\
\emph{Produced\_by} & \emph{Item} $~\xrightarrow{\text{produced\_by}}~$ \emph{Brand} & $0.21\pm0.41$ & $0.17\pm0.38$ & $0.52\pm0.50$ & $0.83\pm0.38$ \\
\emph{Also\_bought} & \emph{Item} $~\xrightarrow{\text{also\_bought}}~$ \emph{Item} & $57.28\pm39.22$ & $61.35\pm32.99$ & $56.53\pm35.82$ & $73.65\pm30.69$ \\
\emph{Also\_viewed}  & \emph{Item} $~\xrightarrow{\text{also\_viewed}}~$ another \emph{Item} & $0.27\pm1.86$ & $6.29\pm6.17$ & $1.24\pm4.29$ & $12.84\pm8.97$ \\
\emph{Bought\_together} & \emph{Item} $~\xrightarrow{\text{bought\_together}}~$ another \emph{Item} & $0.68\pm0.80$ & $0.69\pm0.90$ & $0.81\pm0.77$ & $0.75\pm0.72$ \\
\hline
\end{tabular}
\caption{Descriptions and statistics of four Amazon e-commerce datasets: \emph{CDs \& Vinyl}, \emph{Clothing}, \emph{Cell Phones} and \emph{Beauty}.}
\label{tab:stats}
\vspace{-25pt}
\end{table*}

The final step is to solve our recommendation problem over the knowledge graph guided by the trained policy network.
Recall that given a user $u$, the goal is to find a set of candidate items $\{i_n\}$ and the corresponding reasoning paths $\{p_n(u, i_n)\}$.
One straightforward way is to sample $n$ paths for each user $u$ according to the policy network $\pi(\cdot|\mathbf{s},\mathbf{\tilde{A}}_u)$.
However, this method cannot guarantee the diversity of paths, because the agent guided by the policy network is likely to repeatedly search the same path with the largest cumulative rewards.
Therefore, we propose to employ beam search guided by the action probability and reward to explore the candidate paths as well as the recommended items for each user.
The process is described as Algorithm \ref{alg:beam}. 
It takes as input the given user $u$, the policy network $\pi(\cdot|\mathbf{s},\mathbf{\tilde{A}}_u)$, horizon $T$, and predefined sampling sizes at each step, denoted by $K_1,\ldots,K_T$.
As output, it delivers a candidate set of $T$-hop paths $\mathcal{P}_T$ for the user with corresponding path generative probabilities $\mathcal{Q}_T$ and path rewards $\mathcal{R}_T$.
Note that each path $p_T(u,i_n)\in\mathcal{P}_T$ ends with an item entity associated with a path generative probability and a path reward.

For the acquired candidate paths, there may exist multiple paths between the user $u$ and item $i_n$.
Thus, for each pair of $(u,i_n)$ in the candidate set, we select the path from $\mathcal{P}_T$ with the highest generative probability based on $\mathcal{Q}_T$
as the one to interpret the reasoning process of why item $i_n$ is recommended to $u$.
Finally, we rank the selected interpretable paths according to the path reward in $\mathcal{R}_T$ and recommend the corresponding items to the user.

\section{Experiments}\label{sec:exp}
In this section, we extensively evaluate the performance of our \emph{PGPR} method on  real-world datasets.
We first introduce the benchmarks for our experiments and the corresponding experimental settings. Then we quantitatively compare the effectiveness of our model with other state-of-the-art approaches, followed by ablation studies to show how parameter variations influence our model.

\subsection{Data Description}
All experiments are conducted on the Amazon e-commerce datasets collection \cite{he2016ups}, consisting of product reviews and meta information from Amazon.com.
The datasets include four categories: \emph{CDs and Vinyl}, \emph{Clothing}, \emph{Cell Phones} and \emph{Beauty}.
Each category is considered as an individual benchmark that constitutes a knowledge graph containing 5 types of entities and 7 types of relations.
The description and statistics of each entity and relation can be found in Table \ref{tab:stats}.
Note that once the type of head entity and relation are provided, the type of tail entity is uniquely determined.
In addition, as shown in Table \ref{tab:stats}, we find that \emph{Mention} and \emph{Described\_by} account for a very large proportion among all relations. 
These two relations are both connected to the \emph{Feature} entity, which may contain redundant and less meaningful words. We thus adopt TF-IDF to eliminate less salient features in the preprocessing stage: For each dataset, we keep the frequency of feature words less than 5,000 with TF-IDF score $> 0.1$. We adopt the same data split rule as in previous work \cite{zhang2017joint}, which randomly sampled 70\% of user purchases as the training data and took the rest 30\% as test. The objective in the KGRE-Rec problem is to recommend items purchased by users in the test set together with reasoning paths for each user--item pair.

\begin{table*}[t]
\small
\begin{tabular*}{\textwidth}{c|p{2.3em}p{2.3em}p{2.3em}p{2.3em}|p{2.3em}p{2.3em}p{2.3em}p{2.3em}|p{2.3em}p{2.3em}p{2.3em}p{2.3em}|p{2.3em}p{2.3em}p{2.3em}p{2.3em}}
\hline
Dataset & \multicolumn{4}{c|}{\textbf{CDs \& Vinyl}} & \multicolumn{4}{c|}{\textbf{Clothing}} & \multicolumn{4}{c|}{\textbf{Cell Phones}} & \multicolumn{4}{c}{\textbf{Beauty}} \\
\hline 
Measures ($\%$) & NDCG  & Recall & HR    & Prec. & NDCG  & Recall & HR   & Prec. & NDCG  & Recall & HR    & Prec. & NDCG  & Recall & HR     & Prec. \\
\hline 
BPR            & 2.009 & 2.679 & 8.554  & 1.085 & 0.601 & 1.046 & 1.767 & 0.185 & 1.998 & 3.258 & 5.273  & 0.595 & 2.753 & 4.241  & 8.241  & 1.143\\
BPR-HFT        & 2.661 & 3.570 & 9.926  & 1.268 & 1.067 & 1.819 & 2.872 & 0.297 & 3.151 & 5.307 & 8.125  & 0.860 & 2.934 & 4.459  & 8.268  & 1.132\\
VBPR           & 0.631 & 0.845 & 2.930  & 0.328 & 0.560 & 0.968 & 1.557 & 0.166 & 1.797 & 3.489 & 5.002  & 0.507 & 1.901 & 2.786  & 5.961  & 0.902\\
TransRec       & 3.372 & 5.283 & 11.956 & 1.837 & 1.245 & 2.078 & 3.116 & 0.312 & 3.361 & 6.279 & 8.725  & 0.962 & 3.218 & 4.853  & 0.867  & 1.285\\
DeepCoNN       & 4.218 & 6.001 & 13.857 & 1.681 & 1.310 & 2.332 & 3.286 & 0.229 & 3.636 & 6.353 & 9.913  & 0.999 & 3.359 & 5.429  & 9.807  & 1.200\\
CKE            & 4.620 & 6.483 & 14.541 & 1.779 & 1.502 & 2.509 & 4.275 & 0.388 & 3.995 & 7.005 & 10.809 & 1.070 & 3.717 & 5.938  & 11.043 & 1.371\\
JRL            & 5.378$^*$ & 7.545$^*$ & 16.774$^*$ & 2.085$^*$ & 1.735$^*$ & 2.989$^*$ & 4.634$^*$ & 0.442$^*$ & 4.364$^*$ & 7.510$^*$ & 10.940$^*$ & 1.096$^*$ & 4.396$^*$ & 6.949$^*$ & 12.776$^*$ & 1.546$^*$ \\
PGPR (Ours)  & \textbf{5.590} & \textbf{7.569} & \textbf{16.886} & \textbf{2.157} & \textbf{2.858} & \textbf{4.834} & \textbf{7.020} & \textbf{0.728} & \textbf{5.042} & \textbf{8.416} & \textbf{11.904} & \textbf{1.274} & \textbf{5.449} & \textbf{8.324} & \textbf{14.401} & \textbf{1.707} \\
\hline
\end{tabular*}
\caption{Overall recommendation effectiveness of our method compared to other baselines on four Amazon datasets. The results are reported in percentage (\%) and are calculated based on the top-10 predictions in the test set. The best results are highlighted in bold and the best baseline results are marked with a star ($*$).}
\label{tab:eval}
\vspace{-15pt}
\end{table*}

\subsection{Experimental Setup}
\paragraph{Baselines \& Metrics}
We compare our results against previous state-of-the-art methods.
\textbf{BPR} \cite{rendle2009bpr} is a Bayesian personalized ranking model that learns latent embeddings of users and items.   %
\textbf{BPR-HFT} \cite{mcauley2013hidden} is a Hidden Factors and Topics (HFT) model that incorporates topic distributions to learn latent factors from reviews of users or items. %
\textbf{VBPR} \cite{he2016vbpr} is the Visual Bayesian Personalized Ranking method that builds upon the BPR model but incorporates visual product knowledge. %
\textbf{TransRec} \cite{he2017translation} invokes translation-based embeddings for sequential recommendation. It learns to map both user and item representations in a shared embedding space through personalized translation vectors. %
\textbf{DeepCoNN} or Deep Cooperative Neural Networks \cite{zheng2017joint} are a review-based convolutional recommendation model that learns to encode both users and products with reviews assisting in rating prediction.
\textbf{CKE} or Collaborative Knowledge base Embedding \cite{zhang2016collaborative} is a modern neural recommender system based on a joint model integrating matrix factorization and heterogeneous data formats, including textual contents, visual information and a structural knowledge base to infer the top-$N$ recommendations results. %
\textbf{JRL} \cite{zhang2017joint} is a start-of-the-art joint representation learning model for top-$N$ recommendation that utilizes multimodal information including images, text and ratings into a neural network.
Note that we did not include \cite{wang2018explainable} as a baseline because we are unable to enumerate all the possible paths between user--item pairs due to the large scale of our datasets.

All models are evaluated in terms of four representative top-$N$ recommendation measures: \textbf{Normalized Discounted Cumulative Gain (NDCG)}, \textbf{Recall}, \textbf{Hit Ratio (HR)} and \textbf{Precision (Prec.)}.
These ranking metrics are computed based on the top-10 predictions for every user in the test set. 

\vspace{-5pt}
\paragraph{Implementation Details}
The default parameter settings across all experiments are as follows.
For the KGRE-Rec problem, we set the maximum path length to 3 based on the assumption that shorter paths are more reliable for users to interpret the reasons of recommendation.
For models' latent representations, the embeddings of all entities and relations are trained based on the 1-hop scoring function defined in Equation \ref{eq:transe}, and the embedding size is set to 100.
On the RL side, the history vector $\mathbf{h_t}$ is represented by the concatenation of embeddings of $e_{t-1}$ and $r_t$, so the state vector $\mathbf{s_t}=(\mathbf{u},\mathbf{e_t},\mathbf{e_{t-1}},\mathbf{r_t})$ is of size 400.
The maximum size of the pruned action space is set to 250, i.e., there are at most 250 actions for any state.
To encourage the diversity of paths, we further adopt action dropout on the pruned action space with a rate of $0.5$.
The discount factor $\gamma$ is $0.99$.
For the policy/value network, $\mathbf{W}_1\in\mathds{R}^{400\times512}$, $\mathbf{W}_2\in\mathds{R}^{512\times256}$, $\mathbf{W}_\text{p}\in\mathds{R}^{256\times250}$ and $\mathbf{W}_\text{v}\in\mathds{R}^{256\times 1}$.
For all four datasets, our model is trained for 50 epochs using Adam optimization. We set a learning rate of $0.001$ and a batch size of 64 for the \emph{CDs \& Vinyl} dataset, and a learning rate of $0.0001$ and batch size of 32 for the other datasets. The weight of the entropy loss is $0.001$.
In the path reasoning phase, we set the sampling sizes at each step to $K_1=20, K_2=10, K_3=1$ for \emph{CDs \& Vinyl}, and $K_1=25, K_2=5, K_3=1$ for the other three datasets.

\subsection{Quantitative Analysis}
In this experiment, we quantitatively evaluate the performance of our model on the recommendation problem compared to other baselines on all four Amazon datasets. We follow the default setting as described in the previous section. 

The results are reported in Table \ref{tab:eval}.
Overall, our \emph{PGPR} method consistently outperforms all other baselines on all datasets in terms of NDCG, Hit Rate, Recall and Precision. For example, it obtains a 3.94\% NDCG improvement over the best baseline (JRL) on the \emph{CDs \& Vinyl} dataset, a significant improvement of 64.73\% on \emph{Clothing}, 15.53\% on \emph{Cell Phone}, and 23.95\% on \emph{Beauty}.
Similar trends can be observed for Recall, Hit Rate and Precision on all datasets. This shows that searching for reasoning paths over the knowledge graph provides substantial benefits for product recommendation and hence---even in the absence of interpretability concerns---is a promising technique for recommendation over graphs.
Our path reasoning process is guided by the learned policy network that incorporates rich heterogeneous information from knowledge graphs and captures multi-hop interactions among entities (e.g., user mentions feature, feature is described by item, item belongs to category, etc.).
This also largely contributes to the promising recommendation performance of our method.
Besides, we notice that directly applying TransE for recommendation \cite{ai2018learning} slightly outperforms ours. It can be regarded as a single-hop latent matching method, but the post-hoc explanations do not necessarily reflect the true reason of generating a recommendation. In contrast, our methods generate recommendations through an explicit path reasoning process over knowledge graphs, so that the explanations directly reflect how the decisions are generated, which makes the system transparent.

\begin{table}[t]
	\small
	\centering
	\begin{tabular}{c|ccc}
		\hline
		Dataset & \# Valid Paths/User & \# Items/User & \# Paths/Item \\
		\hline
		CDs \& Vinyl & $173.38\pm 29.63$ & $120.53\pm 25.04$ & $1.44\pm 1.12$ \\
		Clothing & $60.78\pm 7.00$ & $37.21\pm 7.23$ & $1.63\pm 1.25$ \\
		Cell Phone & $117.22\pm13.12$ & $57.99\pm 14.29$ & $2.02\pm 2.34$ \\
		Beauty & $59.95\pm 6.28$ & $36.91\pm 7.24$ & $1.62\pm 1.25$ \\
		\hline
	\end{tabular}
	\caption{Results of number of valid paths per user, number of unique items per user and number of paths per item.}
	\label{tab:valid_path}
	\vspace{-25pt}
\end{table}

\begin{figure*}[t]
	\centering
	\includegraphics[height=1.2in]{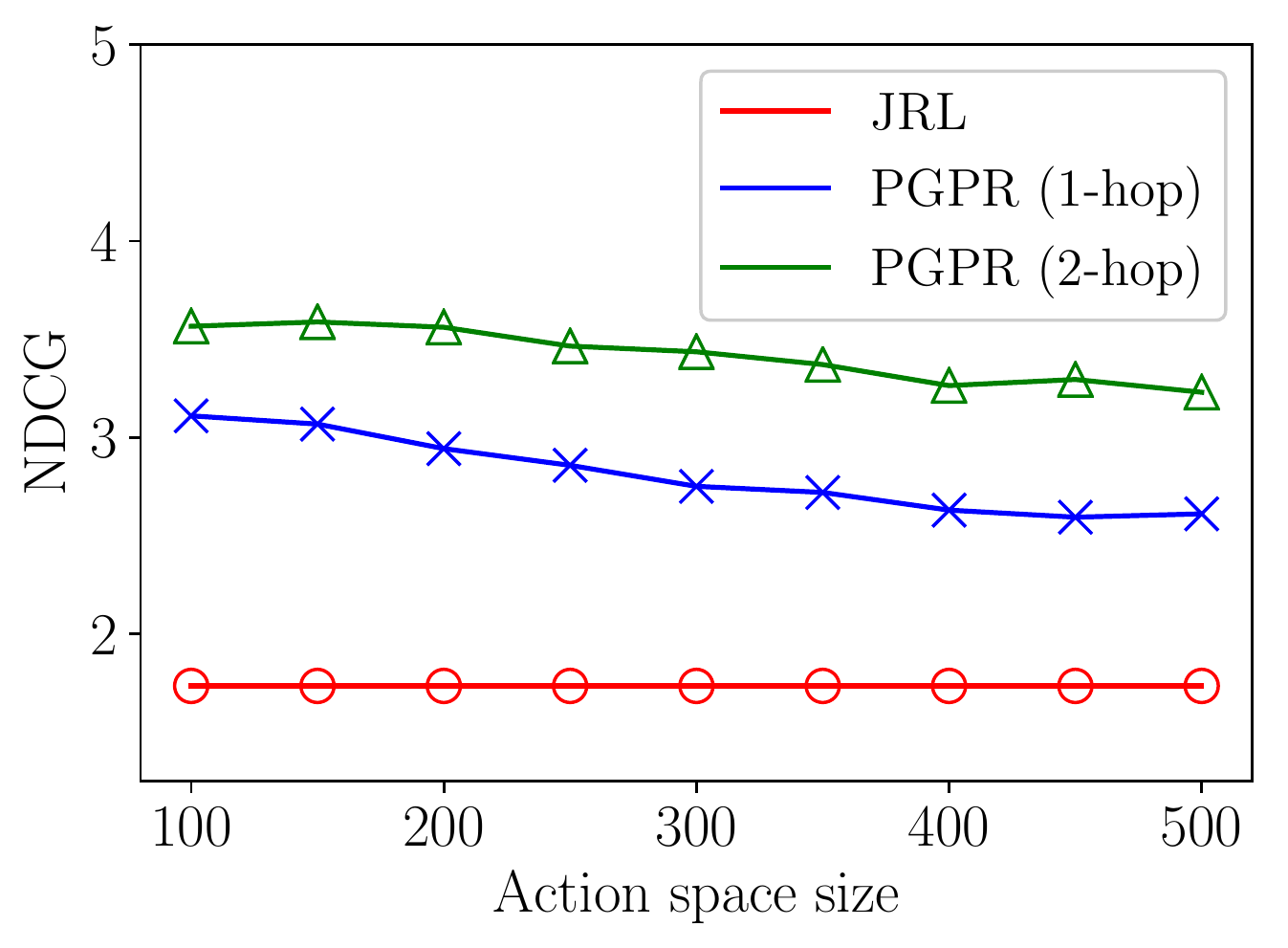}
	\includegraphics[height=1.2in]{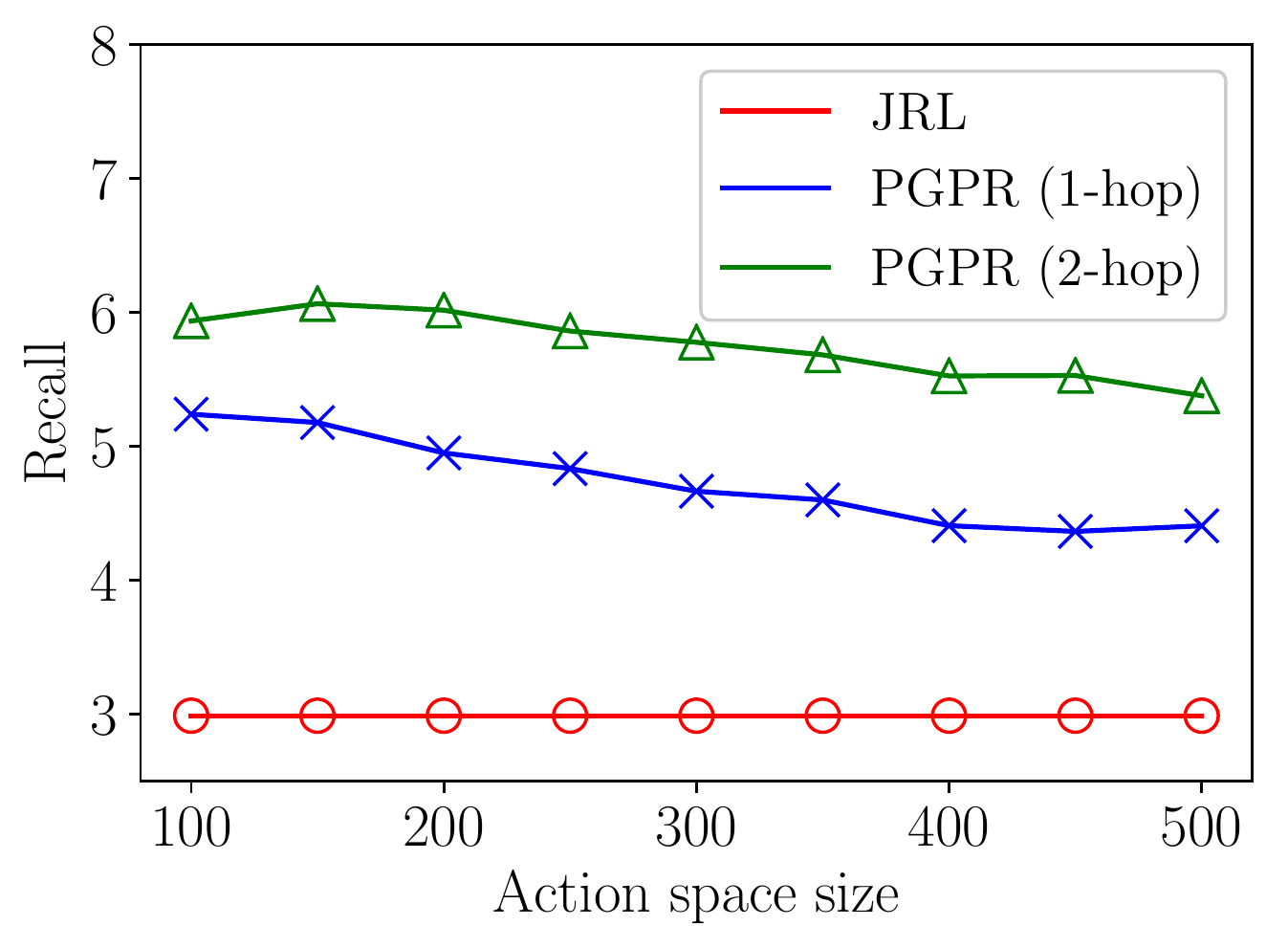}
	\includegraphics[height=1.2in]{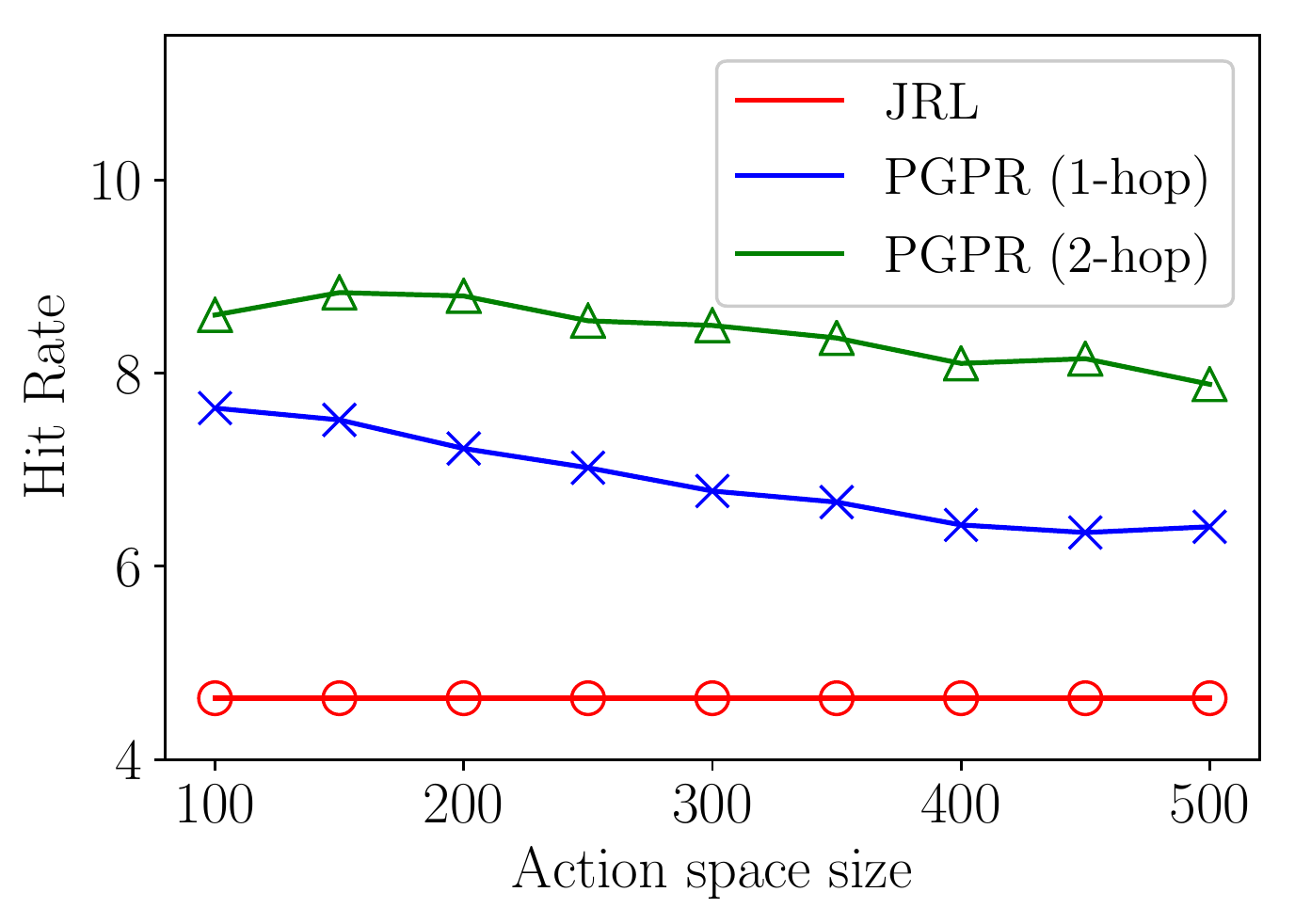}
	\includegraphics[height=1.2in]{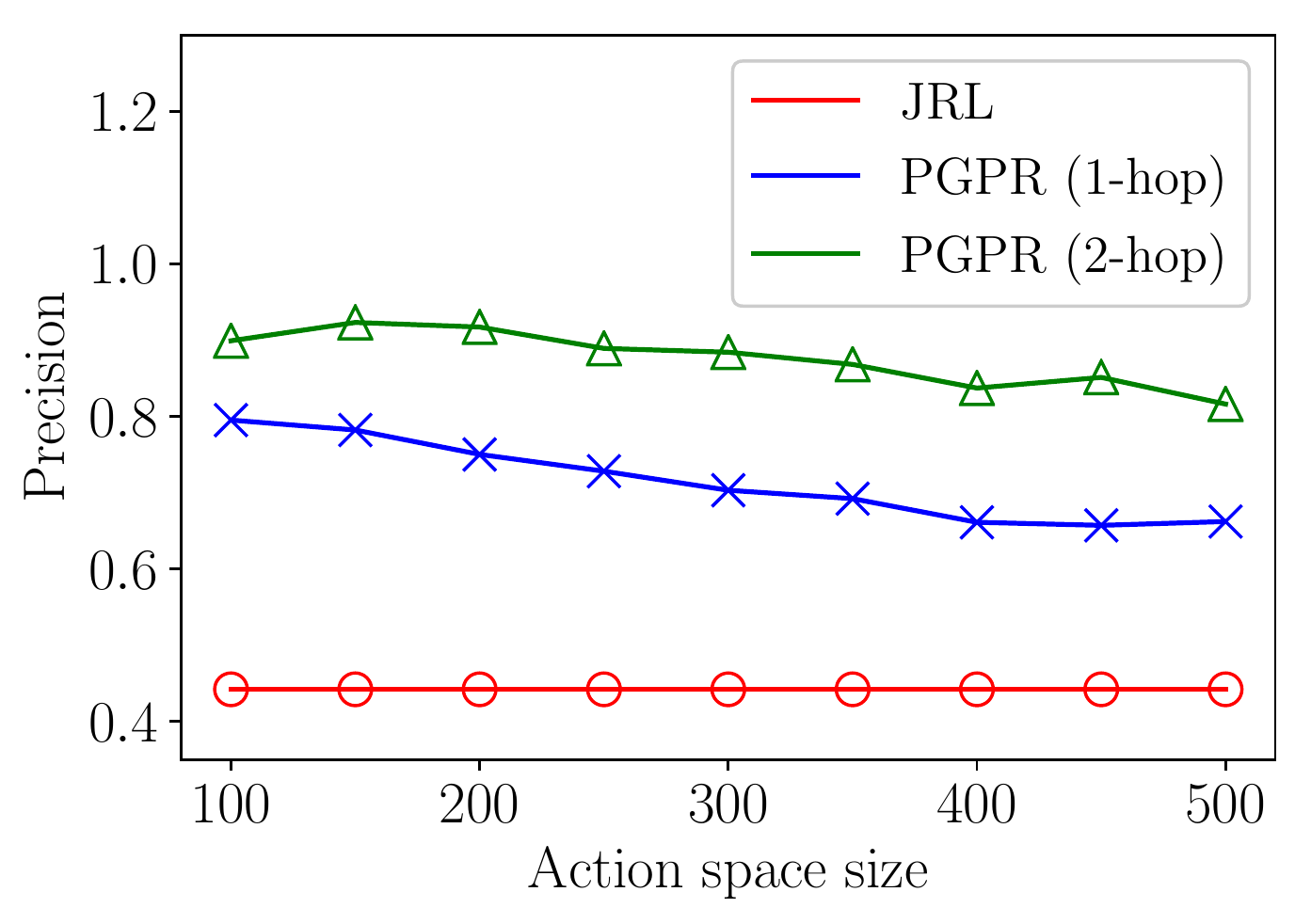} \\
	\hspace{0.2in} (a) NDCG \hspace{1.1in} (b) Recall \hspace{1.1in} (c) Hit Rate \hspace{1.2in} (d) Precision 
	\vspace{-0.1in}
	\caption{Recommendation effectiveness of our model under different sizes of pruned action spaces on the \emph{Clothing} dataset. The results using multi-hop scoring function are also reported.}
	\label{fig:cloth_acts}
	\vspace{-10pt}
\end{figure*}

\begin{figure*}[t]
	\centering
	\includegraphics[height=1.2in]{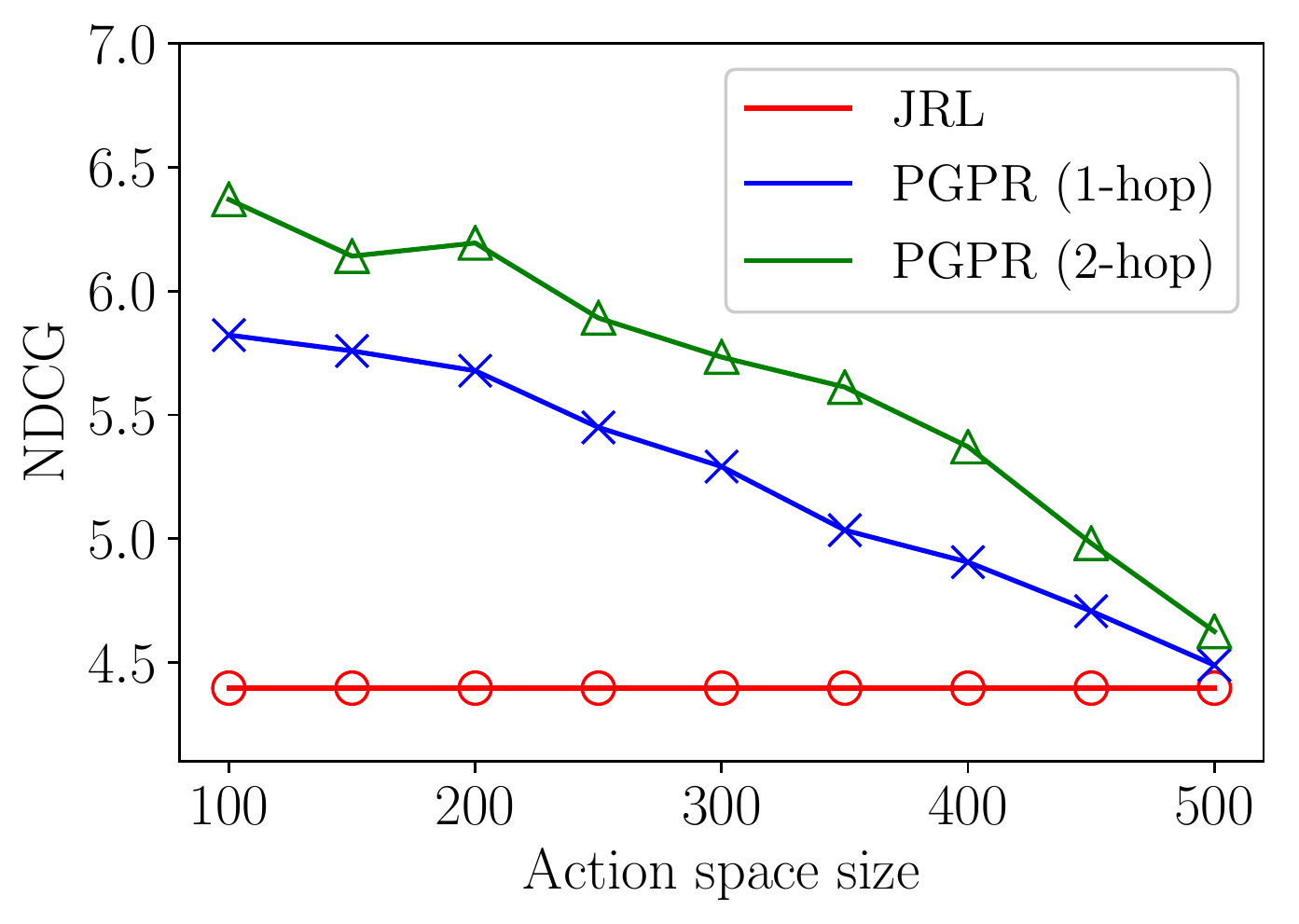}
	\includegraphics[height=1.2in]{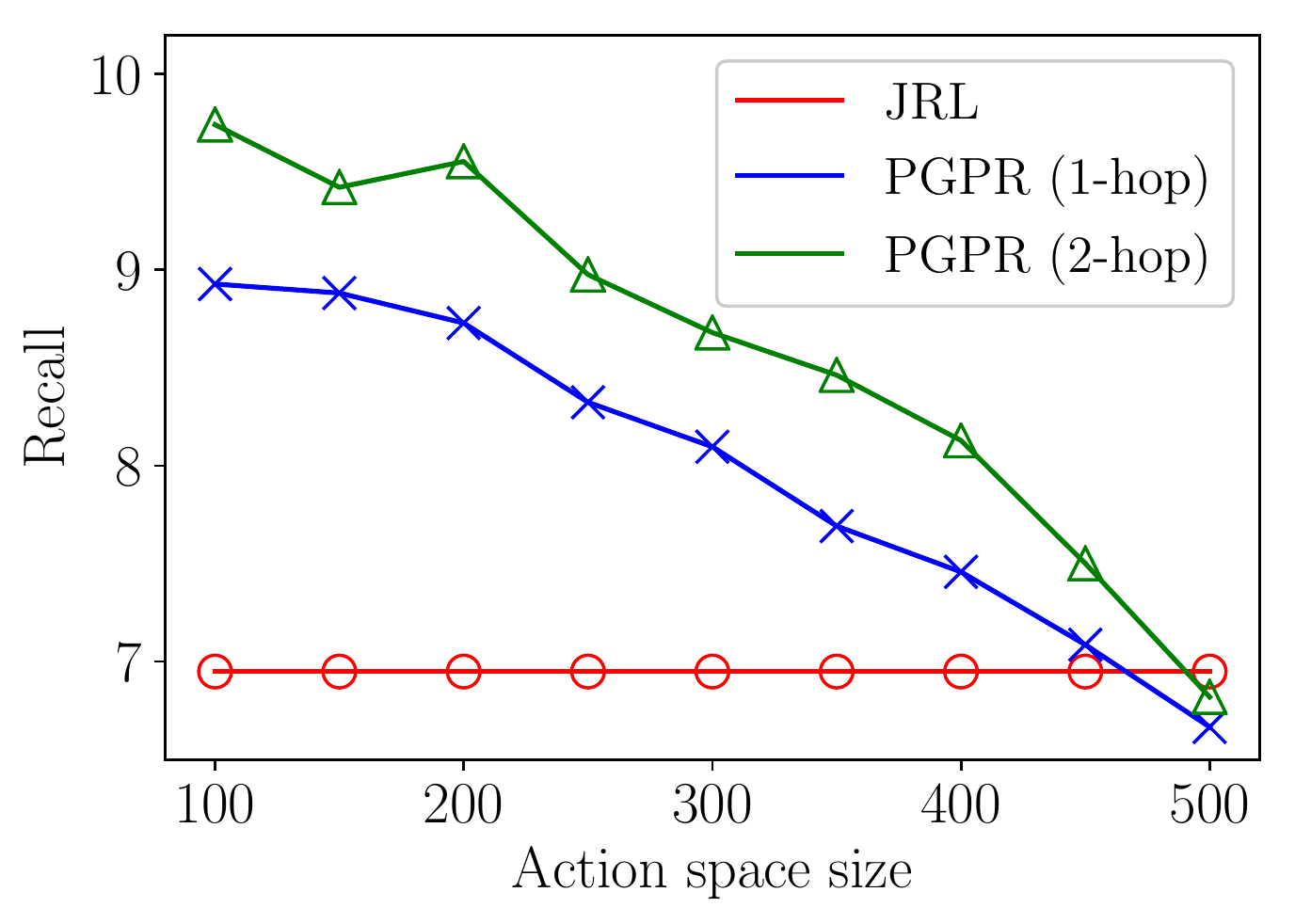}
	\includegraphics[height=1.2in]{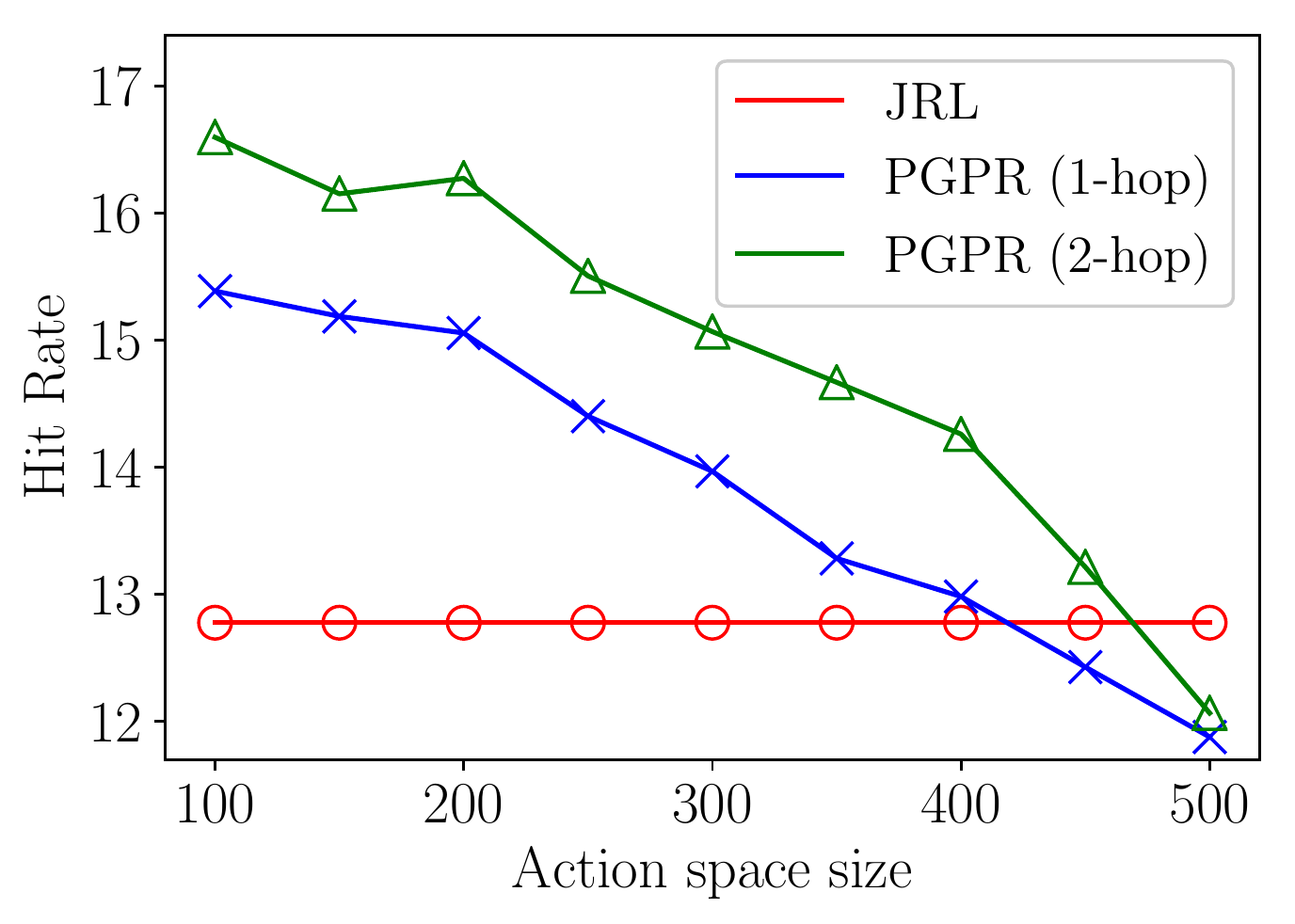}
	\includegraphics[height=1.2in]{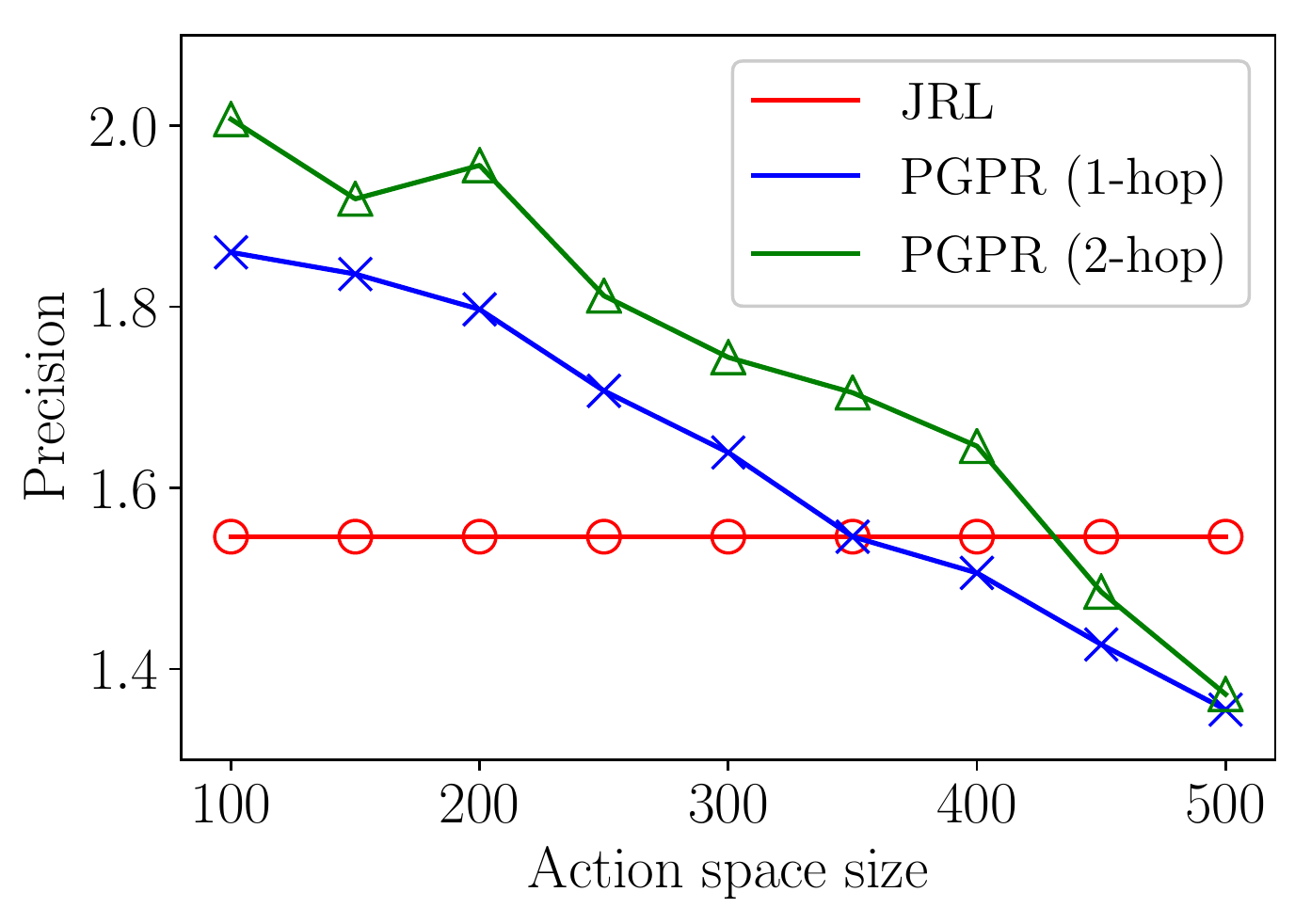} \\
	\hspace{0.2in} (a) NDCG \hspace{1.1in} (b) Recall \hspace{1.1in} (c) Hit Rate \hspace{1.2in} (d) Precision 
	\vspace{-0.1in}
	\caption{Recommendation effectiveness of our model under different sizes of pruned action spaces on the \emph{Beauty} dataset. The results using multi-hop scoring function are also reported.}
	\label{fig:beauty_acts}
	\vspace{-5pt}
\end{figure*}

Furthermore, we examine the efficiency of our method in finding valid reasoning paths.
A path is deemed \emph{valid} if it starts from a user and ends at an item entity within three hops (i.e., at most four entities in a path).
As shown in Table \ref{tab:valid_path}, we respectively report the average number of valid paths per user, the average number of unique items per user, and the average number of supportive paths per item.
We observe two interesting facts. 
First, the success rate of our method to find valid paths is around 50\%, which is calculated as the number of valid paths out of all sampled paths (200 paths for \emph{CDs \& Vinyl} dataset and 125 for others). Especially for the \emph{Cell Phone} dataset, almost all paths are valid.
Considering that the number of all possible paths from each user to items is very large and the difficulty of our recommendation problem is particularly high, these results suggest that our method performs very well in regard to path finding properties.
Second, each recommended item is associated with around 1.6 reasoning paths.
This implies that there are multiple reasoning paths that can serve as supportive evidence for each recommendation. One could hence consider providing more than one of these if users request further details.

\subsection{Influence of Action Pruning Strategy}\label{sec:act}
In this experiment, we evaluate how the performance of our model varies with different sizes of pruned action spaces.

Recall that conditioned on the starting user, the action space is pruned according to the scoring function defined in Equation \ref{eq:score}, where actions with larger scores are more likely to be preserved.
In other words, larger action spaces contain more actions that are less relevant to the user.
This experiment aims to show whether larger action spaces are helpful in exploring more reasoning paths to find potential items.
We experiment on two selected datasets, \emph{Beauty} and \emph{Clothing}, and follow the default setting from the previous section, except that the size of the pruned action space is varied from 100 to 500 with a step size of 50.
The results on two datasets are plotted in Figures \ref{fig:cloth_acts} and \ref{fig:beauty_acts}, respectively.
The best baseline method JRL is also reported in the figures for comparison. Its performance does not depend on the action space.

There are two interesting observations in the results.
First, our model outperforms JRL under most choices of pruned action space sizes. 
Take the \emph{Clothing} dataset as an example. As shown in Figure \ref{fig:cloth_acts}, for any metric among NDCG, Recall, Hit Rate and Precision, the blue curve of our method is consistently above the red curve of JRL by a large margin for all sizes ranging from 100 to 500.
The results further demonstrate the effectiveness of our method compared to other baselines.
Second, the performance of our model is slightly influenced by the size of the pruned action space. 
As shown in both figures, the common trend is that a smaller pruned action space leads to better performance. This means that the scoring function is a good indicator for filtering proper actions conditioned on the starting user. Another possible reason is that larger action spaces require more exploration in RL, but for fair comparison, we set the same parameters such as learning rate and training steps across all different choices of action space, which may lead to suboptimal solutions in  cases of larger action spaces.

\begin{table}[t]
	\centering
	\small
	\begin{tabular}{p{3em}|p{2.em}p{2.em}p{2.em}p{2.em}|p{2.em}p{2.em}p{2.em}p{2.em}}
		\hline
		Dataset & \multicolumn{4}{c|}{\textbf{Clothing}} & \multicolumn{4}{c}{\textbf{Beauty}} \\
		\hline
		Sizes & NDCG & Recall & HR & Prec. & NDCG & Recall & HR & Prec. \\
		\hline
		$25, 5, 1$  & \underline{2.858} & \underline{4.834} & \underline{7.020} & \underline{0.728} & \underline{5.449} & \underline{8.324} & \underline{14.401} & \underline{1.707} \\
		$20, 6, 1$  & 2.918 & 4.943 & 7.217 & 0.749 & 5.555 & 8.470 & 14.611 & 1.749 \\
		$20, 3, 2$  & 2.538 & 4.230 & 6.177 & 0.636 & 4.596 & 6.773 & 12.130 & 1.381 \\
		$15, 8, 1$  & 2.988 & 5.074 & 7.352 & 0.767 & 5.749	& 8.882 & 15.268 & 1.848 \\
		$15, 4, 2$  & 2.605 & 4.348 & 6.354 & 0.654 & 4.829 & 7.138 & 12.687 & 1.458 \\
		$12, 10, 1$ & 3.051 & 5.207 & 7.591 & 0.791 & 5.863 & 9.108 & 15.599 & 1.905 \\
		$12, 5, 2$  & 2.700 & 4.525 & 6.575 & 0.679 & 4.968 & 7.365 & 13.168 & 1.519 \\
		$10, 12, 1$ & \textbf{3.081} & \textbf{5.271} & \textbf{7.673} & \textbf{0.797} & \textbf{5.926} & \textbf{9.166} & \textbf{15.667} & \textbf{1.920} \\
		$10, 6, 2$  & 2.728 & 4.583 & 6.733 & 0.693 & 5.067 & 7.554 & 13.423 & 1.559 \\
		\hline
	\end{tabular}
	\caption{Influence of sampling sizes at each level on the recommendation quality. The best results are highlighted in bold and the results under the default setting are underlined. All numbers in the table are given in percentage (\%).}
	\label{tab:sample_size}
	\vspace{-23pt}
\end{table}

\subsection{Multi-Hop Scoring Function}
Besides action pruning and the reward definition, the scoring function is also used as a part of the objective function in training the knowledge graph representation.
By default, we employ a 1-hop scoring function for representation learning.
In this experiment, we explore whether multi-hop scoring functions can further improve the recommendation performance of our method.

In particular, the 2-hop scoring function from Equation \ref{eq:score} is:
$f(e_0,e_2|\tilde{r}_{2,2})=\langle \mathbf{e_0}+\mathbf{r_1}+\mathbf{r_2}, \mathbf{e_2} \rangle+b_{e_2}$
for any valid 2-hop pattern $\tilde{r}_{2,2}=\{r_1,r_2\}$ between entities $e_0$ and $e_2$.
This function is plugged into the objective function in Equation \ref{eq:embed_obj} for training entity and relation embeddings.
All further settings are adopted from the previous action space experiment. 
We also plot the results in Figures \ref{fig:cloth_acts} and \ref{fig:beauty_acts} with an additional green curve representing our new model trained by the 2-hop scoring function.
Surprisingly, we find that our 2-hop PGPR method further outperforms the default model (blue curve).
This improvement mainly stems from the effectiveness of the multi-hop scoring function, which captures interactions between entities with longer distance. For example, if a user purchases an item and the item belongs to a category, the 2-hop scoring function enhances the relevance between the \emph{User} entity and the \emph{Category} entity through the 2-hop pattern \{\emph{Purchase}, \emph{Belong\_to}\}. 

\begin{table}[t]
	\small
	\centering
	\begin{tabular}{p{3em}|p{2.em}p{2.em}p{2.em}p{2.em}|p{2.em}p{2.em}p{2.em}p{2.em}}
		\hline
		Dataset & \multicolumn{4}{c|}{\textbf{Clothing}} & \multicolumn{4}{c}{\textbf{Beauty}} \\
		\hline
		History & NDCG & Recall & HR & Prec. & NDCG & Recall & HR & Prec. \\
		\hline
		0-step & 1.972 & 3.117 & 4.492 & 0.462 & 3.236 & 4.407 & 8.026 & 0.888 \\
		1-step & 2.858 & 4.834 & 7.020 & 0.728 & 5.449 & 8.324 & 14.401 & 1.707 \\
		2-step & 2.786 & 4.702 & 6.865 & 0.710 & 5.342 & 8.181 & 14.168 & 1.669 \\
		\hline
	\end{tabular}
	\caption{Results for different history representations of state. All numbers in the table are given in percentage (\%).}
	\label{tab:exp_state}
	\vspace{-10pt}
\end{table}

\begin{figure}[t]
    \vspace{-10pt}
	\includegraphics[width=0.43\textwidth]{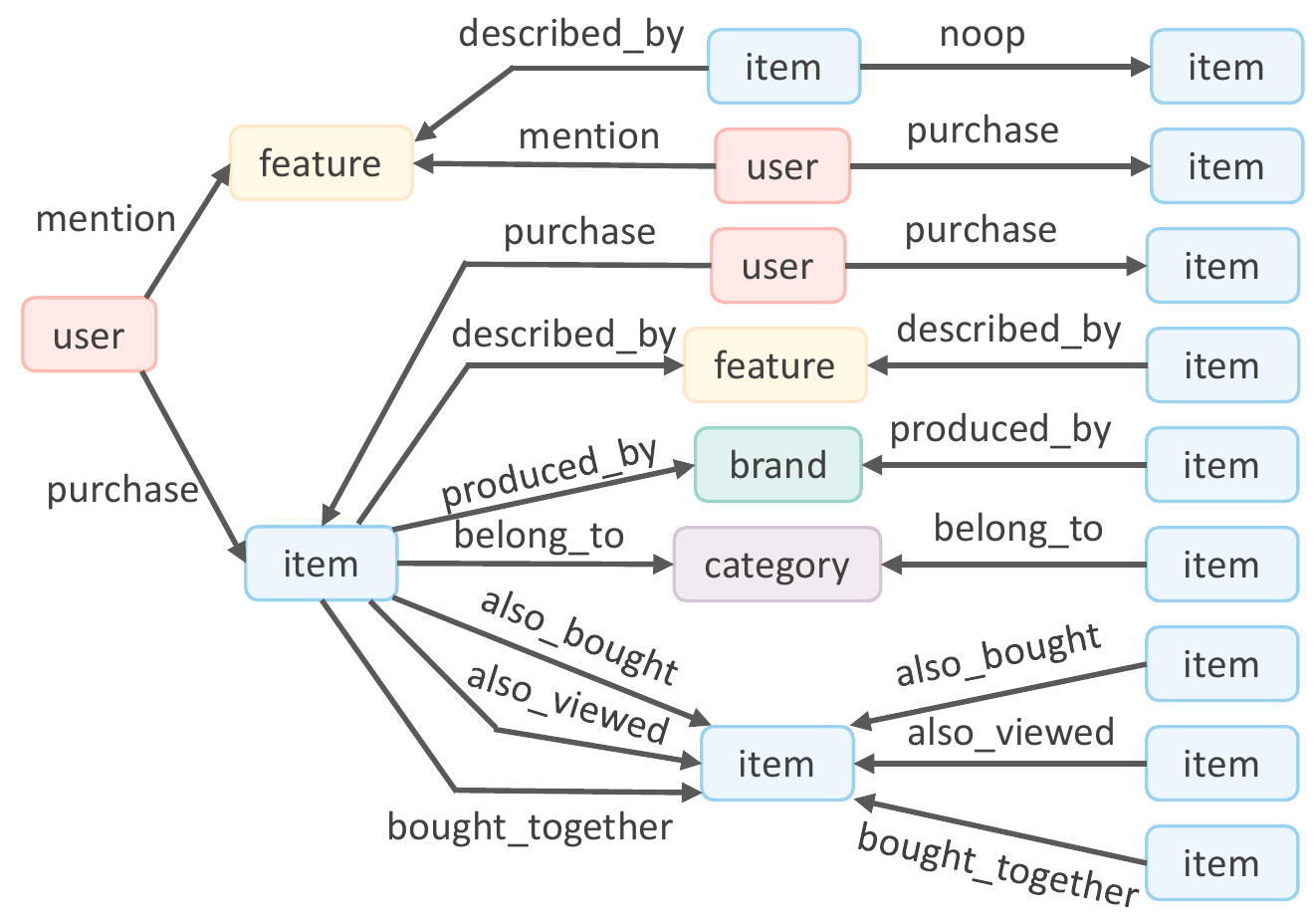}
	\vspace{-10pt}
	\caption{All 3-hop path patterns found in the results.}
	\label{fig:path_pattern}
	\vspace{-10pt}
\end{figure}

\subsection{Sampling Size in Path Reasoning}
In this experiment, we study how the sampling size for path reasoning influences the recommendation performance of our method.

We carefully design 9 different combinations of sampling sizes given a path length of 3. 
As listed in the first column of Table \ref{tab:sample_size}, each tuple $(K_1, K_2, K_3)$ means that the we sample top $K_t$ actions at step $t$ as described in Algorithm \ref{alg:beam}.
For fair comparison, the total number of sampling paths ($=K_1\times K_2\times K_3$) is fixed to 120 (except for the first case).
We experiment on the \emph{Clothing} and \emph{Beauty} datasets and follow the default settings of other parameters.
The recommendation results in terms of NDCG, Recall, Hit Rate and Precision are reported in Table \ref{tab:sample_size}.
Interestingly, we observe that the first two levels of sampling sizes play a more significant role in finding good paths.
For example, in the cases of $(25,5,1), (20,6,1), (15,8,1),(12,10,1),(10,12,1)$, our model performs much better than in the rest of cases.
One explanation is that the first two selections of actions largely determine what kinds of items can be reached. After the first two steps are determined, the policy network tends to converge to selecting the optimal action leading to a good item.
On the other hand, our model is quite stable if the sample sizes at the first two levels are large, which offers a good guidance for parameter tuning.

\subsection{History Representations}
Finally, we examine how different representations of state history influence our method.
We consider three alternatives for $\mathbf{h}_t$: no history (0-step), last entity $e_{t-1}$ with relation $r_t$ (1-step), and last two entities $e_{t-2},e_{t-1}$ with relations $r_{t-1},r_{t}$ (2-step).
Other settings are the same as in the previous experiments.
As shown in Table \ref{tab:exp_state}, we find that the worst results are obtained in the 0-step case, which suggests that a state representation without history cannot provide sufficient information for the RL agent to learn a good policy.
Apart from this, the performance of using 2-step history is slightly worse than that of 1-step history. One possible reason is that additional history information is redundant and even misleads the algorithm in the decision-making process.

\section{Case Study on Path Reasoning}
To intuitively understand how our model interprets the recommendation, we give a case study here based on the results generated in the previous experiments.
We first study the path patterns discovered by our model during the reasoning process, followed by various cases for recommendation.

\paragraph{Path patterns}
For a fixed path length of 3, we find that our method managed to discover 15 different path patterns, which are plotted in an aggregated form in Figure \ref{fig:path_pattern}.
Interestingly, the pattern $\big\{$\textit{user $\xrightarrow{purchase}$ item $\xleftarrow{purchase}$ user $\xrightarrow{purchase}$ item}$\big\}$ is one kind of collaborative filtering.

\paragraph{Case study}
\begin{figure}[t]
\includegraphics[width=0.46\textwidth]{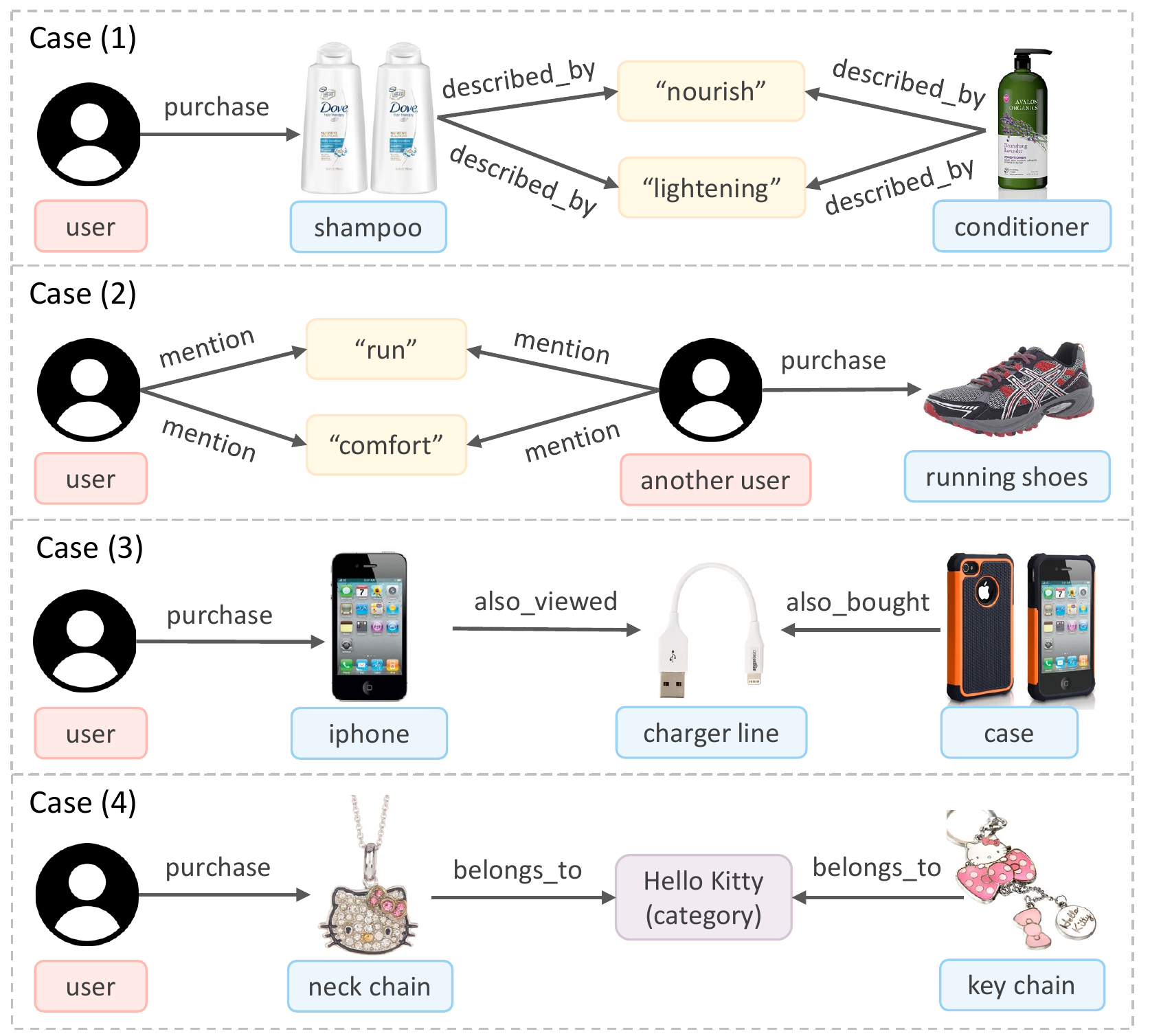}
\vspace{-10pt}
\caption{Real cases of recommendation reasoning paths.}
\label{fig:case_study}
\vspace{-15pt}
\end{figure}
As shown in Figure \ref{fig:case_study}, we provide several real-world examples of the reasoning paths generated by our method to illustrate how to interpret recommendations through paths. 

The first example (Case 1) comes from the \emph{Beauty} dataset, where a user purchased an item ``shampoo'' that was described by two feature words ``nourish'' and ``lightening''. Meanwhile, another item ``conditioner'' also contained these two features in some review. Therefore, our model recommended ``conditioner'' to this user. %
In the second example (Case 2), there are two users who both mentioned the feature words ``run'' and ``comfort'' in their reviews, so our method made a decision based on the purchase history of one user,  by recommending the item ``running shoes'' purchased by the user to the other user. 
In the third example (Case 3), a user bought an item  ``iPhone'' and also viewed another item ``charger line''. Considering that other users who purchased ``phone case'' would also buy ``charger line'', our method accordingly recommended ``iPhone case'' to the user. 
The last example (Case 4) depicts that one user purchased an item ``neck chain'', which belonged to the category of ``Hello Kitty''. Thus, our method recommended the user another item ``key chain'' that was also in the same category ``Hello Kitty''.
We conclude that our PGPR method not only achieves promising recommendation results, but also is able to efficiently find diverse reasoning paths for the recommendations.

\section{Conclusions and Future Work}
We believe that future intelligent agents should have the ability to perform explicit reasoning over knowledge for decision making. In this paper, we propose RL-based reasoning over knowledge graphs for recommendation with interpretation. To achieve this, we develop a method called Policy-Guided Path Reasoning (PGPR). Based on our proposed soft reward strategy, user-conditional action pruning strategy, and a multi-hop scoring approach, our RL-based PGPR algorithm is not only capable of reaching outstanding recommendation results, but also exposes its reasoning procedure for explainability. We conduct extensive experiments to verify the performance of our approach compared with several state-of-art baselines. 
It should be noted that our PGPR approach is a flexible graph reasoning framework and can be extended to many other graph-based tasks such as product search and social recommendation, which will be explored in the future. We can also extend our PGPR approach to model time-evolving graphs so as to provide dynamic decision support.

\bibliographystyle{ACM-Reference-Format}
\bibliography{paper}
\balance

\end{document}